\documentclass[twocolumn,3p,times,procedia]{elsarticle}

\usepackage{graphicx}
\usepackage{balance}
\usepackage{array,threeparttable}
\usepackage{url}
\usepackage{amssymb}
\usepackage{url}
\usepackage{amsmath,amssymb,amsfonts}
\usepackage{graphicx}
\usepackage{textcomp}
\usepackage{xcolor}
\usepackage{array}
\usepackage{url}
\usepackage{color,framed}
\usepackage{color,xcolor}
\usepackage{threeparttable}
\usepackage{multirow}
\usepackage{longtable}
\usepackage{algorithm}  
\usepackage{setspace}
\usepackage{array}
\usepackage{booktabs}
\usepackage{amsmath}
\usepackage{amssymb}
\usepackage{listings,xcolor}
\usepackage{courier}
\usepackage{caption}   
\usepackage{graphicx}  
\usepackage{caption}
\usepackage{subfigure} 
\usepackage{pdflscape}
\usepackage{bbding}
\usepackage{adjustbox}
\usepackage{enumerate}
\usepackage{blkarray}
\usepackage{comment}
\usepackage{blkarray}
\usepackage{algpseudocode,float}
\usepackage{lipsum}
\usepackage{changepage}
\usepackage{colortbl}
\usepackage{geometry}
\newcommand{\tabincell}[2]{\begin{tabular}{@{}#1@{}}#2\end{tabular}}
\newcommand{\reffig}[1]{Fig.\ref{#1}}
\newcommand{\refargo}[1]{Algorithm.\ref{#1}}
\newcommand{\refeqs}[1]{Eq.\ref{#1}}

\makeatletter
\newenvironment{breakablealgorithm}
  {
   \begin{center}
     \refstepcounter{algorithm}
     \hrule height.8pt depth0pt \kern2pt
     \renewcommand{\caption}[2][\relax]{
       {\raggedright\textbf{\ALG@name~\thealgorithm} ##2\par}%
       \ifx\relax##1\relax 
         \addcontentsline{loa}{algorithm}{\protect\numberline{\thealgorithm}##2}%
       \else 
         \addcontentsline{loa}{algorithm}{\protect\numberline{\thealgorithm}##1}%
       \fi
       \kern2pt\hrule\kern2pt
     }
  }{
     \kern2pt\hrule\relax
   \end{center}
  }
\makeatother
\usepackage{amssymb}
\usepackage{amsmath}
\usepackage{multicol}
\usepackage{algorithm}
\usepackage{algpseudocode}

\usepackage[figuresright]{rotating}
\usepackage{bm}
\usepackage{caption}
\usepackage{titlesec}
\titleformat*{\section}{\small \bf}
\titleformat*{\subsection}{\small \em}
\titleformat*{\subsubsection}{\small \em}

\usepackage{fancyhdr}
\fancyheadoffset[RO,EL]{0pt}
\usepackage{geometry}
\begin{document}\small
\begin{frontmatter}




\title{
\begin{flushleft}
{\LARGE Mitigating Blockchain extractable value (BEV) threats by Distributed Transaction Sequencing in Blockchains}
\end{flushleft}
}
 %

\author[]{ \leftline {Xiongfei Zhao$^a$, Hou-Wan Long $^b$, Zhengzhe Li $^c$, Jiangchuan Liu $^d$, Yain-Whar Si $^*$$^a$}}

\address{ 

  \leftline {$^a$Department of Computer and Information Science, University of Macau, Macau}

  \leftline {$^b$Department of Statistics, The Chinese University of Hong Kong, Hong Kong}

  \leftline {$^c$Department of Mathematics, New York University, USA}

  \leftline {$^d$School of Computing Science, Simon Fraser University, Canada}

}

\fntext[]{Xiongfei Zhao has more than 18 years of experience working in the technology department at large banks. He received his BSc degree in Computer Science and Technology from Inner Mongolia University. His research interests include Blockchain, Computational Intelligence and Financial Technology. (email:yb97480@um.edu.mo).}

\fntext[]{Hou-Wan Long is an undergraduate student in the Department of Statistics at The Chinese University of Hong Kong, China, where he is currently pursuing his studies with a major focus on Risk Management Science. (email:1155190681@link.cuhk.edu.hk).}

\fntext[]{Zhengzhe Li is a senior student in the Department of Mathematics at New York University. He was a summer intern in the Financial Modeling Group in BlackRock in 2023. (email:zl3440@nyu.edu).}

\fntext[]{Jiangchuan Liu is currently a Full Professor (with University Professorship) in the School of Computing Science at Simon Fraser University, British Columbia, Canada. He is a Fellow of The Canadian Academy of Engineering, an IEEE Fellow, and an NSERC E.W.R. Steacie Memorial Fellow. (email:jcliu@sfu.ca).}

\cortext[]{Yain-Whar Si (Corresponding author) will handle correspondence at all stages of refereeing and publication, also post publication (email:fstasp@um.edu.mo).}

\begin{abstract}

\textcolor[RGB]{0,0,0}{The rapid growth of Blockchain and Decentralized Finance (DeFi) has introduced new challenges and vulnerabilities that threaten the integrity and efficiency of the ecosystem. This study identifies critical issues such as Transaction Order Dependence (TOD), Blockchain Extractable Value (BEV), and Transaction Importance Diversity (TID), which collectively undermine the fairness and security of DeFi systems. BEV-related activities, including Sandwich attacks, Liquidations, and Transaction Replay, have emerged as significant threats, collectively generating \$540.54 million in losses over 32 months across 11,289 addresses, involving 49,691 cryptocurrencies and 60,830 on-chain markets. These attacks exploit transaction mechanics to manipulate asset prices and extract value at the expense of other participants, with Sandwich attacks being particularly impactful. Additionally, the growing adoption of Blockchain in traditional finance highlights the challenge of TID, where high transaction volumes can strain systems and compromise time-sensitive operations. To address these pressing issues, we propose a novel Distributed Transaction Sequencing Strategy (DTSS), which combines forking mechanisms and the Analytic Hierarchy Process (AHP) to enforce fair and transparent transaction ordering in a decentralized manner. Our approach is further enhanced by an optimization framework and the introduction of the Normalized Allocation Disparity Metric (NADM), which ensures optimal parameter selection for transaction prioritization. Experimental evaluations demonstrate that DTSS effectively mitigates BEV risks, enhances transaction fairness, and significantly improves the security and transparency of DeFi ecosystems. This work is essential for protecting the future of decentralized finance and promoting its integration into global financial systems.}

\end{abstract}


\end{frontmatter}

\section{Introduction}


Blockchains comprise a network of globally distributed P2P nodes. A financial transaction is deemed confirmed by the network when it is included in at least one Blockchain block, which possesses the most ``Proof of Work''. Blockchain nodes store unconfirmed transactions in the mempool. Due to the nature of Blockchain mining, which consolidates multiple transactions into a single block based on transaction fees, the process deviates from conventional centralized systems that process transactions chronologically.

\textcolor[RGB]{0,0,0}{Decentralized Finance (DeFi), built on Blockchain technology, has witnessed substantial growth, exceeding 90B USD in locked value and handling millions of daily transactions. The transparency provided by permissionless Blockchains renders DeFi an appealing alternative to traditional finance. DeFi network users transmit transactions to their P2P neighbors until they reach a miner. Simultaneously, miners are integral to block creation and possess the exclusive authority to determine transaction order within the blocks, an informational asymmetry that can be leveraged for financial gain \cite{zhou2021high} \cite{10521704}. }

\textcolor[RGB]{0,0,0}{Qin et al. \cite{qin2022quantifying} conducted a comprehensive analysis to characterize market manipulation behaviors associated with Transaction Ordering Dependence (TOD) issues, where the sequence in which transactions are processed affects their outcomes. The study specifically focused on Blockchain Extractable Value (BEV) activities, which refer to profits obtained by manipulating transaction orders within a blockchain, carried out by predatory traders through transaction fee manipulation, as well as miners taking advantage of their mining positions.} Additionally, the researchers categorized the risk of Blockchain Extractable Value (BEV) by assessing the financial impact resulting from various manipulative practices, including Sandwich attacks, Liquidations, and Transaction Replay. By quantifying the USD extracted through these manipulative techniques, indicating a total profit of 540.54 million USD generated over a period of 32 months. This profit was distributed across 11,289 addresses, involving a vast array of 49,691 cryptocurrencies and encompassing 60,830 on-chain markets.

\begin{figure}[H]
    \makeatletter
    \def\@captype{figure} 
    \makeatother
    \centering
    \includegraphics[width=0.46 \textwidth]{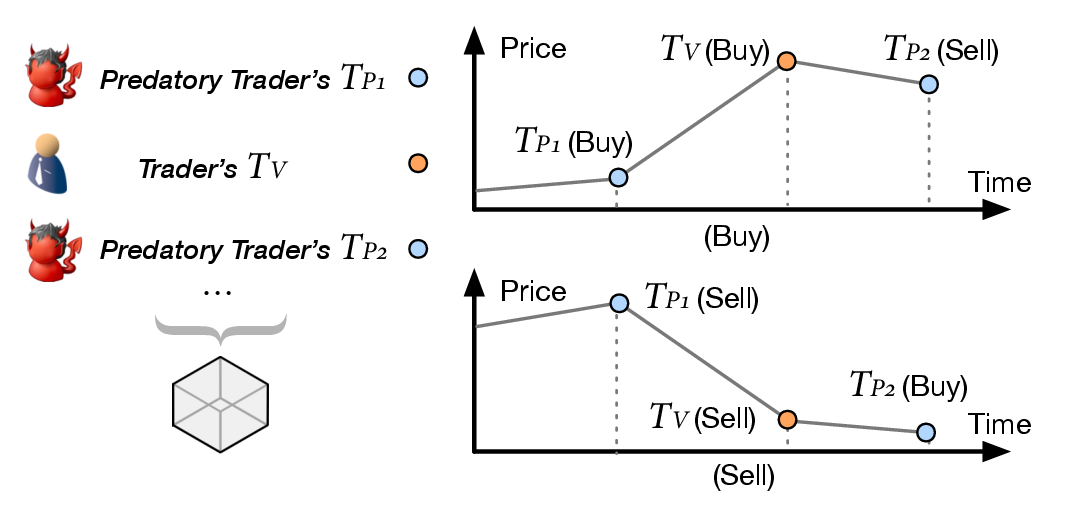}
    \caption{Illustration of Sandwich attacks and two scenarios: ``buy'' and ``sell''. In scenario (Buy), predatory trader buys the same asset prior to $T_{V}$ affecting the market price, enabling predatory trader to sell the asset at a higher price following the transaction. In scenario (Sell), predatory trader sells the same asset before $T_{V}$ impacts the market price, allowing predatory trader to repurchase the asset at a lower price after the transaction.}
    \label{SandwichAttack}
\end{figure}

\textcolor{black}{The Sandwich Attack, a form of Blockchain Extractable Value (BEV), constitutes a potential risk to transactional processes within the Decentralized Finance (DeFi) ecosystem.} As depicted in \reffig{SandwichAttack}, wherein a trader strategically inserts their own transactions between a victim transaction. The purpose of this attack is to exploit the anticipated price movement of an asset following the execution of a ``large'' pending transaction (referred to as $T_{V}$). The predatory trader, who can be a miner or trader, monitors the peer-to-peer (P2P) network for pending transactions. If the market price of the asset is expected to increase or decrease after $T_{V}$ is executed, the predatory trader proceeds with the Sandwich attack.

The Sandwich attack is segregated into two stages: The predatory trader, denoted as $P$, initiates transaction $T_{P1}$, which aims to preemptively take advantage of the price movement caused by $T_{V}$. This is accomplished by purchasing or selling the same asset before $T_{V}$ influences the market price. Subsequently, $P$ executes transaction $T_{P2}$ to counteract the effects of $T_{P1}$ and close the trading position established in the previous step. This ensures that $P$ can benefit from the price change caused by $T_{V}$. 

\textcolor[RGB]{0,0,0}{As the DeFi landscape continues to evolve, traditional financial institutions are also adapting and innovating to remain competitive in this rapidly changing landscape. However, the growing popularity of Blockchain technology presents another challenge: Transaction Importance Diversity (TID) \cite{10.1145/3284028.3284035}, which refers to the varying levels of importance and urgency among different types of blockchain transactions. Financial industries such as banks, exchanges, insurance companies, and payment agencies use Blockchain as distributed ledgers that record important business processes. When Blockchain serves as a financial business infrastructure, it allows for the processing of immutable logs of payments, securities transactions, wire transfers, statements, etc., thereby increasing transparency, efficiency, and trust. However, high-volume scenarios can strain the system because a large influx of non-time-sensitive transactions such as statements can hinder the efficiency of time-sensitive transactions such as security trading transactions. Addressing this ETS challenge requires acknowledging the various business importance and service requirements of Blockchain transactions.}


To tackle the challenges posed by Transaction Order Dependence (TOD) and Blockchain extractable Value (BEV), as well as to support Transaction Importance Diversity (TID), we propose a novel mechanism called Distributed Transaction Sequencing Strategy (DTSS). DTSS leverages forking to ensure miners' adherence to predefined transaction ordering rules in a distributed manner. Under DTSS, transactions are assigned varying weight values based on their attributes. For instance, traders strategically inserting their own transactions between victim transactions with a later initiation time would carry a higher weight value. Alternatively, transactions of lower importance may be assigned higher weight values. The incorporation of higher weight transactions leads to the creation of a larger block, denoted as Block A'. Conversely, miners following DTSS and incorporating transactions with lower weight values result in the formation of a smaller block, referred to as Block A. When Block A' is created and presented to the network, it competes with Block A to become part of the longest chain. 

DTSS utilizes the Analytic Hierarchy Process (AHP) to assign weights to transaction attributes, thereby highlighting the varying priorities of distinct transaction types within a versatile Blockchain platform. The AHP's pairwise comparison tables, a fundamental component of DTSS methodology, facilitate the analysis of relationships between discrete and continuous variables such as Transaction Type, Amount, Fee, and Initiation Time. By employing optimization algorithms, we identify the optimal settings for constructing the AHP comparison table, contributing to informed decision-making and efficient block size determination. Furthermore, by harnessing the potential of forking threats, we cultivate a robust incentive for miners to adhere to the correct transaction prioritization.

The contributions of our proposal are as follows: 

\begin{itemize}
    
\item \textcolor{black}{The relationship between forking and block size has been investigated. Based on our analysis results, we observe that smaller block sizes in Blockchain networks yield lower forking probabilities, incentivizing miners to favor conditions conducive to smaller blocks for profit optimization.}

\item Propose a novel mechanism, named DTSS, which dynamically adapts block size according to transaction attributes, thereby enforcing transaction execution order to tackle the challenges posed by Transaction Order Dependence (TOD) and Blockchain extractable Value (BEV), as well as to effectively accommodate Transaction Importance Diversity (TID).
    
\item Introduce an optimization framework, along with a Normalized Allocation Disparity Metric (NADM)  benchmark, to ascertain the optimal parameter set for DTSS. This approach aids in calculating appropriate weights for different transaction attributes, thereby ensuring optimal transaction prioritization.

\end{itemize}

\textcolor{black}{This paper is organized as follows. Section 2 summarizes the related work. Section 3 introduces Distributed Transaction Sequencing Strategy (DTSS). Section 4 describes the experiment settings and experimental results. In Section 5, an analysis of the transaction ordering compliance result and forking rate of blocks is presented. The paper is concluded in Section 6 with future work.}

\section{Background and Related Work}





In traditional financial markets, user transactions are typically processed and organized by a trusted and regulated intermediary based on the order in which they are received. However, in the context of a Blockchain, user transactions within a block operate on a different set of principles. Specifically, transactions within a block are treated as atomic and deterministic. This means that multiple transactions within the same block are consolidated into a single action and executed in an ``all-or-nothing'' manner. Moreover, within a given Blockchain state, the execution of a user's transaction is deterministic \cite{qin2022quantifying}. These transaction processing rules in the Blockchain differ from those in traditional financial markets and have a significant impact on the dynamics of value extraction within the Blockchain ecosystem. 

Parties involved in a transaction can attempt to manipulate the order of transactions in order to prioritize adversarial transactions within the Blockchain state. By doing so, they aim to maximize their own revenue in a competitive manner, distinct from the traditional financial market dynamics \cite{auer2022miners}. \textcolor{black}{The subsequent sections delve into the intricacies of Blockchain transaction handling, starting with Transaction Ordering Dependence (TOD) which highlights the sequential dependencies of transactions. Building on this, the concept of Blockchain Extractable Value (BEV) illustrates potential profits from transaction order manipulation for all network participants. This leads to a discussion on specific BEV-related attacks, including Sandwich Attacks and Liquidations, showing the real-world implications of such manipulations. This exploration culminates with the examination of Transaction Importance Diversity (TID), which highlights the complexities and potential integrity threats within transaction handling in Blockchain networks. These interconnected topics provide a comprehensive view of the vulnerabilities and challenges inherent to Blockchain transaction processing.}


\subsection{Transaction Ordering Dependence (TOD)}

Transaction ordering dependence in Blockchain refers to how subsequent transactions rely on the order in which previous transactions are processed and added to the Blockchain. Transactions are grouped into blocks and added to the Blockchain sequentially based on the consensus mechanism. The order of transactions within a block is significant because later transactions may depend on the outcomes of earlier ones. This is especially important in smart contracts, where multiple transactions are interconnected. Manipulating the order of transactions within a block or across different blocks can have serious implications. Malicious actors may attempt to reorder transactions to their advantage, which can lead to unfair advantages, double-spending attacks, or other forms of manipulation that compromise the integrity and trustworthiness of the Blockchain system \cite{7546538} \cite{10.1145/2976749.2978309}.

\subsection{Blockchain Extractable Value (BEV)}

With TOD as a foundational concept, Miner Extractable Value (MEV) was subsequently introduced by Daian et al. in 2020 \cite{daian2020flash}, which denotes the potential profit that miners can attain by strategically manipulating the selection and ordering of transactions within a Blockchain network. MEV encompasses the economic advantage that miners can leverage through their ability to influence the incorporation and order of transactions in the Blockchain. It serves as a comprehensive term for capturing the financial benefits that miners can gain from these actions. Qin et al. (2022) \cite{qin2022quantifying} expand upon this notion by highlighting that not only miners but also non-mining traders possess the ability to impact the incorporation and order of transactions by adjusting factors such as transaction fees. As a result, Qin et al. broaden the scope of MEV to encompass the behavior of any participant within the Blockchain ecosystem who seeks financial gains by controlling transaction order. To reflect this broader perspective, they introduce the concept of Blockchain Extractable Value (BEV).

Within the realm of MEV, Qin et al. \cite{qin2022quantifying} introduce several prominent attack types, namely Sandwich attacks, clearing, arbitrage, and trade replay. Each attack exhibits distinct characteristics and holds significant implications, as elucidated in the following descriptions.

\textbf{\textit{Sandwich Attacks}}: Sandwich attacks are a predatory trading strategy where a trader wraps a victim transaction between two adversarial transactions. The attacker, who can be a miner or trader, exploits market movements by strategically adding buy or sell transactions just before large pending transactions, profiting at the expense of other market participants. The attack comprises two steps: (i) issuing transaction to front-run victim transaction by buying or selling the same asset before victim transaction impacts the market price, and (ii) issuing transaction to back-run victim transaction and closing the trading position. This profit is obtained not only at the expense of other market players but also causes genuine transactions to be processed more slowly, imposing an ``invisible tax'' on other market participants.

\textbf{\textit{Liquidation}}: Liquidations play a pivotal role in the Blockchain-based decentralized finance (DeFi) ecosystem, and two widely adopted mechanisms have emerged. The first mechanism, known as fixed spread liquidation, provides a solution for cases where borrowers default on their debt obligations. In such instances, a liquidator is granted the opportunity to purchase the collateral at a predetermined discount. By acquiring the collateral at a reduced price, the liquidator effectively mitigates the risks associated with the defaulted loan, safeguarding the lending platform. The second mechanism, auction liquidation, involves the initiation of an auction process by a liquidator, which operates within a predefined duration. During this auction, competing liquidators submit bids based on the lowest price they are willing to pay for the collateral. By maximizing the amount recovered, the auction liquidation mechanism contributes to the overall stability and integrity of DeFi lending protocols. However, opportunistic traders or miners could actively monitor pending liquidation transactions and adeptly execute their own transactions ahead of the liquidation process. This proactive approach enables these traders to capitalize on the liquidation opportunity, employing strategic maneuvering to gain a competitive edge over other market participants.

\textbf{\textit{Arbitrage}}: Arbitrage is a strategic practice that involves the simultaneous purchase and sale of assets in different markets to capitalize on price discrepancies, thereby fostering market efficiency. In the realm of Blockchain, arbitrage leverages variations in prices, liquidity, and trading volumes across decentralized exchanges (DEXs) and networks. By exploiting these disparities, Blockchain arbitrageurs aim to secure assets at lower prices and sell them at higher prices, aligning market prices and enhancing liquidity. 


\textbf{\textit{Transaction Replay}}: Transaction replay refers to the process of replicating and re-executing an unconfirmed transaction from one network onto another. This method can be exploited by adversaries to extract value by copying the execution logic of a victim transaction and diverting the revenue to their controlled account. The steps involved in Transaction Replay include observing a victim transaction on the network layer, constructing replay transactions to mimic the execution logic, locally validating the replay transactions, and, if profitable, attempting to front-run the victim transaction. This method allows adversaries to exploit vulnerabilities and potentially disrupt the intended outcome of transactions, enabling them to gain an advantage over other participants in the Blockchain network. 

\textbf{\textit{Clogging}}: A Clogging refers to a malicious act aimed at occupying block space on a Blockchain network, thereby hindering the timely incorporation of other legitimate transactions. In order to execute a Clogging, the attacker seeks out opportunities, such as Liquidations or gambling scenarios, where immediate extraction of monetary value is not possible. The attacker then broadcasts transactions with high fees and computational requirements to congest the pending transaction queue.

\subsection{Transaction Importance Diversity (TID)}

Goel et al. \cite{10.1145/3284028.3284035} highlighted the application of Blockchain technology in sectors such as banking, insurance, and supply chain logistics, using the supply chain network as an illustrative example. Within this framework, organizations partake in transactions, contribute to the consensus mechanism, and manage distributed ledgers within the same Blockchain network, thereby creating a cohesive ecosystem. Each organization may assume distinct roles within a given business process. In the context of a supply chain network, a multitude of entities, including manufacturers, suppliers, retailers, logistics providers, warehouses, and financiers, may be involved. In such a scenario, the ledger meticulously logs information related to shipments, status updates, purchase orders, and invoices, along with actions undertaken by various entities. However, in high-volume situations, the system may encounter challenges, as a surge in record-keeping transactions on the Blockchain can delay the execution of critical business operations. This underscores the idea that transactions vary in their business importance and place diverse service demands on the Blockchain infrastructure.



\textcolor{black}{Confronting the ongoing challenges of BEV and TID, several strategies have been put forward. Fair Ordering \cite{10.1007/978-3-030-56877-1_16} and its extension \cite{10.1145/3494105.3526239} introduced by Kelkar et al., both of which focus on achieving order fairness. Zhou et al. \cite{zhou2021a2mm} have proposed an application-specific BEV mitigation strategy using an Automated Market Maker design. Moreover, a weighted fair queuing strategy \cite{10.1145/3284028.3284035} has been suggested to optimize transaction handling based on priority. However, none of these strategies propose a similar approach to DTSS which uniquely utilizes forking as a mechanism to enforce transaction priority rules. Therefore, while we have thoroughly reviewed the existing work, we reiterate that our approach stands distinct in the current landscape of Blockchain transaction handling solutions.}

\subsection{Fair Ordering}

Kelkar et al. \cite{10.1007/978-3-030-56877-1_16} introduce a novel consensus protocol, referred to as \textsl{Aequitas}. Uniquely, \textsl{Aequitas} protocols attain order-fairness alongside consistency and liveness, expanding the capabilities of existing consensus protocols. These protocols employ two fundamental primitives in a black-box manner: FIFO Broadcast (FIFO-BC), an extension of standard reliable broadcast, and Set Byzantine Agreement (Set-BA), achievable from the Byzantine agreement. 

\subsection{Fair Ordering Extension}

Kelkar et al. \cite{10.1145/3494105.3526239} further present an extension of permissioned order fairness to a permissionless setting, focusing on robustness against rapidly churning adversaries. We propose two permissionless fair ordering protocols, $\Pi_{mod}$, and $\Pi_{fairfruit}$, derived from any longest-chain protocol. These protocols utilize a novel structure, named Hammurabi, in which each transaction's ordering is proposed across multiple blocks. The system mines independent transaction lists, ``semantic chains'', from which the final fair ordering is extracted. This approach mitigates the potential manipulation by adversaries, ensuring a fair overall ordering. 

\subsection{Application-Specific BEV Mitigation}

Zhou et al. \cite{zhou2021a2mm} introduce a novel Automated Market Maker (AMM) design, termed as Automated Arbitrage Market Maker (A$^2$MM), which inherently conducts optimal trade routing and best-effort two-point arbitrage among associated AMMs. The A$^2$MM design bolsters Blockchain security by atomically extracting two-point arbitrage MEV from peered AMMs, discouraging competitive network layer bidding. Following the exchange, a trader can promptly initiate an arbitrage, potentially leading to extra financial gains, it precludes any adversary from appropriating the profits derived from arbitrage.

\subsection{Weighted Fair Queueing Strategy}

The weighted fair queuing strategy \cite{10.1145/3284028.3284035} optimizes the handling of diverse transaction types with varying priorities on the Blockchain. Transactions are submitted by clients to a group of Priority Calculators, which assign transaction priority based on pre-established criteria. These assigned transactions are then forwarded to Ordering Service Nodes. These nodes consolidate the diverse priorities assigned by endorsers into a single priority value, following a predetermined consolidation policy. To ensure fair handling of different priority levels, the system maintains multiple transaction queues, each corresponding to a specific priority level. Transactions are then read by a Multi-Queue Block Generator, following a priority-aware block formation policy, resulting in the generation of a block.

\section{Distributed Transaction Sequencing Strategy (DTSS) Overview and its Relationship to Forking}

\textcolor{black}{In traditional Blockchain systems, miners have the freedom to choose transactions from the mempool to include in a block. The order of these transactions within the block is also at the discretion of the miners. This freedom, however, can lead to unfavorable situations such as BEV attacks, where miners or traders take advantage of the transaction order to their benefit, and TID issues, where certain types of transactions (e.g., non-time-sensitive transactions) can hinder the efficiency of other, more important transactions (e.g., time-sensitive transactions).}

\textcolor{black}{In this study, we propose Distributed Transaction Sequencing Strategy (DTSS) as a solution to challenges encountered in Blockchain-based financial systems.  These challenges include Transaction Order Dependence (TOD), Blockchain Extractable Value (BEV), and Transaction Importance Diversity (TID).  Consequently, DTSS seeks to replicate the fair trade execution order intrinsic to conventional finance, thereby maintaining the principles of fairness, efficiency, and transparency when integrating Blockchain technology into traditional financial systems.  DTSS achieves this by employing forking to enforce transaction ordering rules and assigning weight values to transactions based on factors such as initiation time or importance.  This approach results in the generation of diverse block sizes that vie for inclusion in the longest Blockchain.}

\textcolor{black}{To assign weights, DTSS leverages the Analytic Hierarchy Process (AHP) and optimization algorithms, enabling informed decision-making block size determination. By leveraging forking, DTSS incentivizes correct transaction prioritization by miners. These contributions enhance the versatility and efficiency of Blockchain platforms, accommodating diverse transaction importance and addressing sequencing challenges. While forking presents risks, such as divergent transaction histories and manipulation, we see it as an opportunity for innovation and system resilience.}

\textcolor{black}{The principal goal of DTSS is to strategically influence fork occurrences in the Blockchain. By enforcing transaction ordering rules through forking, DTSS creates different block sizes that compete for inclusion in the longest chain. We aim to increase the probability of forks in blocks mined by non-compliant miners by enlarging the block size, while simultaneously reducing the likelihood of forks in blocks mined by compliant miners through block size reduction. Our research process begins by establishing a mathematical relationship between forking and block size.}



\subsection{Forking and Block Size}

\textcolor{black}{Forking, in this context, refers to the creation of an alternative version of the blockchain, leaving two blocks where there was only one before. This often happens when two miners solve the proof-of-work problem and create a new block at roughly the same time. We aim to investigate the correlation between smaller block sizes and lower probabilities of forking within Blockchain networks. We denote a block as $B$ and its size as $Size(B)$. The propagation of a block between two nodes, $N_f$ and $N_t$, is largely driven by the available ``Network Bandwidth (Mbps) Between Each Pair of Regions,'' which we will denote as $Bandwidth(N_f, N_t)$. Additionally, the propagation time of a block from $N_f$ to $N_t$ is also affected by the ``Average Network Delay Between Each Pair of Regions,'' represented as $Delay(N_f, N_t)$. Given these factors, we can construct a formula to estimate the block propagation time, $TP_{n}$, from node $N_{f}$ to node $N_{t}$: }

\begin{footnotesize}
    \begin{equation}
        TP_{n} =  \displaystyle\frac{Size(B)}{Bandwidth(N_{f},N_{t})} + Delay(N_{f},N_{t})
        \label{propagation}
    \end{equation}
\end{footnotesize}

\textcolor{black}{As the new block is mined, it is being propagated to the entire Blockchain network using a P2P mechanism. For a block to be propagated through $i$ nodes, it needs to be propagated $i-1$ times. Therefore we denote the total propagation time among $i$ nodes for a block as $TP_{i}$ (see \refeqs{total_propagation}).}

\begin{footnotesize}
    \begin{equation}
        TP_{i}=\displaystyle\sum_{n=1}^{i-1}TP_{n}
        \label{total_propagation}
    \end{equation}
\end{footnotesize}

\textcolor{black}{In the study conducted by Christian et al. \cite{6688704}, the authors introduced algorithms to compute the probability of the network discovering a block ($P_b$) at any given second. Here, $X_b$ signifies the random variable representing the time difference, measured in seconds, between the discovery of a block and its predecessor. The expression $P_r[X_b < t + 1 | X_b \geq t]$ denotes the conditional probability of the network finding a block in the subsequent second ($t + 1$), provided that a block has not been identified up to the current second ($t$).} 

\vspace{-0.2cm}
\begin{footnotesize}
    \begin{equation}
        \displaystyle
            P_b=P_r\left[X_b < t + 1 \mid X_b \geq t \right] \approx \frac{1}{600}
        \label{BlockMiningProb}
    \end{equation}
\end{footnotesize}

\textcolor{black}{Whether node $j$ has successfully received block $b$ sent by other nodes through block relay protocol at time $t$, can be defined as:} 

\vspace{-0.2cm}
\begin{footnotesize}
    \begin{equation}
        \displaystyle
            I_j(t)=\begin{cases}
                0,& t_j > t\\
                1,& t_j \leq t
            \end{cases}
    \end{equation}
\end{footnotesize}

\textcolor{black}{For a network with $V$ nodes, the number of nodes that have already received propagated block at time $t$ can be defined as follows:} 

\vspace{-0.2cm}
\begin{footnotesize}
  \begin{equation}
    \displaystyle
            I(t)=\sum_{j \in V} I_j(t)
  \end{equation}
\end{footnotesize}

\textcolor{black}{For the number of nodes $I(t)$ that have already received propagated block at time $t$, the total propagation time can be defined as follows:} 

\vspace{-0.2cm}
\begin{footnotesize}
  \begin{equation}
    \displaystyle
            TP(I(t))=\displaystyle\sum_{n=1}^{I(t)-1}TP_{n}
            \label{PropTime}
  \end{equation}
\end{footnotesize}

Based on \refeqs{propagation} and \refeqs{PropTime}, it can be observed that for nodes that receive blocks at time $t$ in a Blockchain network, the total time $TP(I(t))$ transmitted between nodes is influenced by the size of the block. Specifically, when the size of the block is smaller, the total time $TP(I(t))$ required for transmission between nodes is also smaller. This relationship implies that smaller blocks are more likely to be transmitted between nodes faster compared to larger blocks.

\textcolor{black}{Next, the ratio of nodes that have already received propagated block at time $t$ is defined based on the expectation of $I(t)$ as follows:} 

\vspace{-0.2cm}
\begin{footnotesize}
    \begin{equation}
        \displaystyle
        f(t) = \mathbb{E}\left[I(t)\right] \cdot n^{-1}
    \end{equation}
\end{footnotesize}

\textcolor{black}{Nodes that have not yet received the block that is being propagated may generate conflicting blocks \cite{LIU2021335}. By combining the probability of mining a new block with the ratio of nodes that have not received a propagated block, Christian et al. \cite{6688704} derive the probability of forking. In \refeqs{ForkRatio}, $F$ is the number of the conflicting blocks mined when a block is propagating.}

\begin{footnotesize}
    \begin{equation}
        \displaystyle
        P_r\left[F \geq 1 \right] = 1 - (1-P_b)^{\int_0^\infty{(1-f(t))dt}}
        \label{ForkRatio}
    \end{equation}
\end{footnotesize}

When the size of the block is smaller, the block can be transmitted more quickly across the network, reducing the likelihood of delays and ensuring that more nodes receive the block at a given time $t$. Therefore, the ratio of nodes that have received the propagated block $f(t)$ will also be larger. Subsequently, in \refeqs{ForkRatio}, \textcolor{black}{$\int_0^\infty{(1-f(t))dt}$ will be smaller and the number of nodes that have not received the block will decrease. As a result, the probability of the conflict block being mined $P_r\left[F \geq 1 \right]$ is reduced.} This reduction in the probability of a conflict block being mined subsequently results in a lower probability of forking  $P_r$. 

Based on the analysis presented, it can be concluded that smaller blocks are associated with lower probabilities of forking in Blockchain networks. Miners continually seek to maximize their profits by ensuring the successful integration of their blocks into the Blockchain. Understanding that smaller blocks are less susceptible to forking, miners have a compelling incentive to comply with conditions that favor smaller block sizes.

\section{Detailed Implementation of Distributed Transaction Sequencing Strategy (DTSS)}

\begin{figure*}
\centering
\includegraphics[width=0.7\linewidth]{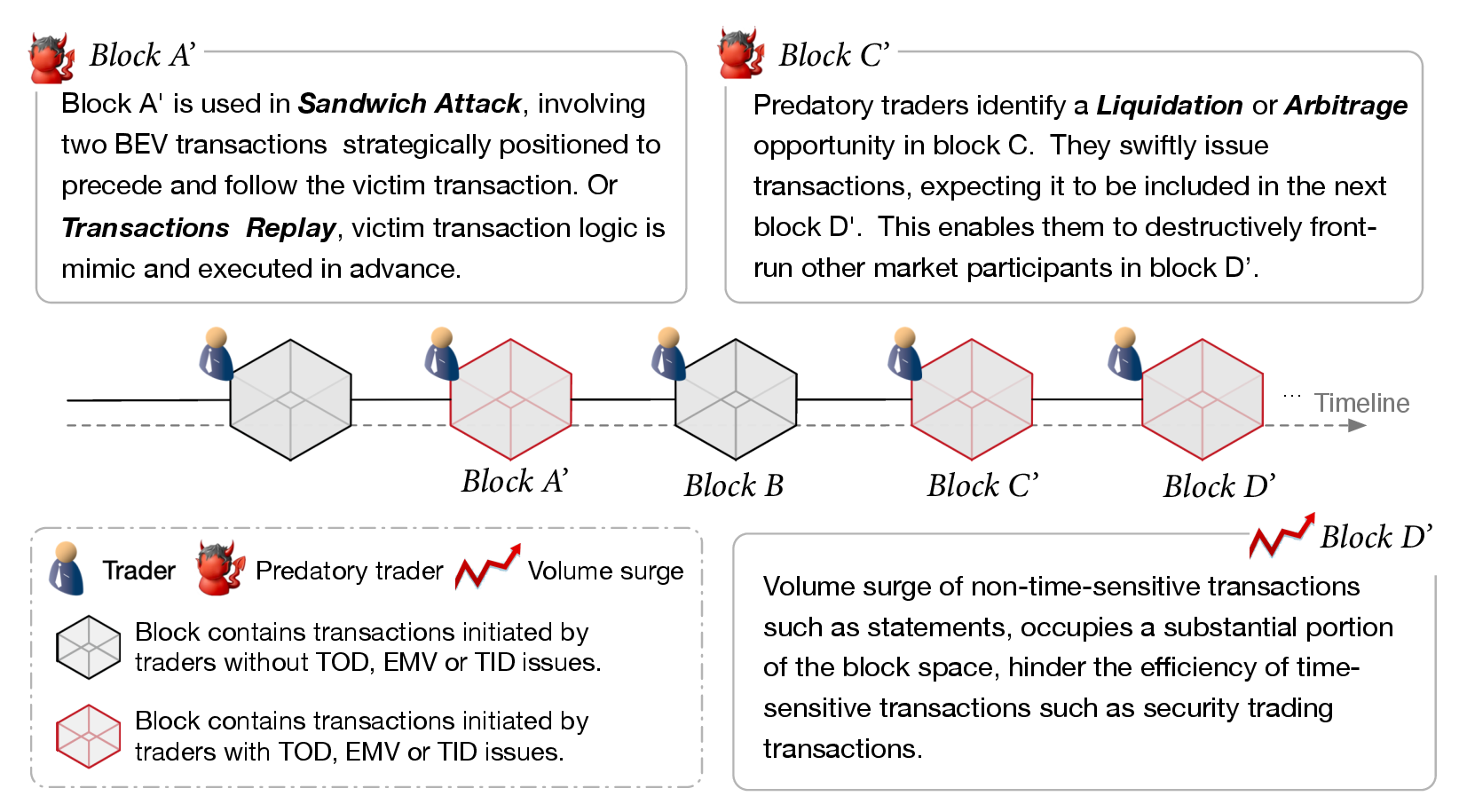}
\caption{Blockchain ecosystem experiences a range of issues related to BEV and TID challenges, including predatory traders' Sandwich and Replay Attacks (Block A'), exploitation of liquidation or arbitrage opportunities (Block C'), and efficiency hindrance from non-time-sensitive transactions (Block D').}
    \label{DTSS0}
\end{figure*}

\textcolor{black}{We will initiate our discussion by presenting a visual representation of BEV and TID issues in Figure \ref{DTSS0}. These complexities lie at the heart of the problems that DTSS is distinctly engineered to address.} The first scenario is a Sandwich attack depicted in block A', where an attacker executes two BEV transactions to front-run and back-run a victim transaction.  Block A' also involves a Transaction Replay attack, where the attacker replicates the victim's transaction to divert profits.  The second scenario involves predatory traders in Block C' identifying opportunities for liquidation or arbitrage in Block B and strategically executing trades to gain an advantage over other participants in Block C'. Lastly, in Block D', non-time-sensitive transactions, like statements, occupying a substantial portion of the block space due to volume surge, hinder the efficiency of time-sensitive transactions, such as security trading.  Blocks encompassing routine transactions are integrally linked with blocks that incorporate transactions demonstrating BEV and TID complexities, collaboratively constituting the longest chain.

\noindent

\textcolor{black}{Moving into the solution aspect of our discourse, we present Distributed Transaction Sequencing Strategy (DTSS). DTSS exploits the intrinsic correlation between block sizes and forking probabilities to steer the incorporation of transactions. This strategy establishes ordering prerequisites for transactions, using the Analytic Hierarchy Process (AHP). In the event miners do not comply with these stipulations, their created block is at risk of being forked and supplanted by a block that aligns with the transaction sequencing guidelines. This mechanism encourages miners to comply with the established transaction ordering prerequisites.}


In the ensuing discussion, we will encompass a comprehensive walkthrough of DTSS's complete process flow (see \reffig{DTSS2}). Within a consortium Blockchain environment, DTSS is embedded into the consensus protocol. Miners are required to adhere to DTSS when selecting transactions to be incorporated into the block. DTSS initially includes two AHP comparison tables that define two priority vectors $V_1$ and $V_2$ for the computation of priority factors. Parameters within the AHP tables are pre-calculated based on the types of financial services processed by the Blockchain network and the processing priority requirements for transactions.

\subsection{Mining stage}

\textcolor{black}{DTSS employs a multi-step process beginning with order transactions in the mempool via an Analytic Hierarchy Process (AHP) table. The Kendall Tau Coefficient calculation is then applied to preserve transaction order within the block. Block size is finally dictated by each transaction's score and leaf space allocation through the Cumulative Distribution Function (CDF). This integrated approach is explained in further detail in the subsequent sections:}

\begin{description}
    
\item[Step1] \textbf{\textsl{Transaction Selection}}

As depicted in \reffig{DTSS2}, in the \textsl{Mining Stage}, \textsl{Step1 Transaction Selection}, miners calculate the scores of all transactions in the mempool according to the AHP table before transaction selection. The score, $S_i$, for transaction $i$ is determined by multiplying the priority vectors $V_1$ and $V_2$ defined in AHP by the attribute matrix of transaction $Ti$. \reffig{DTSS2} illustrates two transaction types. \textcolor{black}{Red dots symbolize BEV transactions, which are strategically crafted to augment the value extracted from the Blockchain. Red dots also represent non-time-sensitive TID transactions, which due to their lower urgency can be executed at a slower pace without negatively impacting the overall Blockchain operations. In contrast, gray dots denote regular or time-sensitive transactions.} 

Under DTSS, by prioritizing transactions in the mempool, we target to inflate the cost for miners to select BEV or TID transactions from mempool, thereby lessening their appeal. However, it is noteworthy that the specific sequence these transactions assumes within a block also carries substantial importance. Despite the utilization of the AHP algorithm to rank transactions in mempool under DTSS, there remains a possibility that BEV transactions could still quantify a high ranking under DTSS. These highly ranked BEV transactions could potentially be incorporated into the block by miners, alongside victim transactions. Therefore, transaction ordering within a block is also instrumental in thwarting BEV attacks and safeguarding dependencies. 

\noindent


\begin{figure*}
\centering
\includegraphics[width=0.7\linewidth]{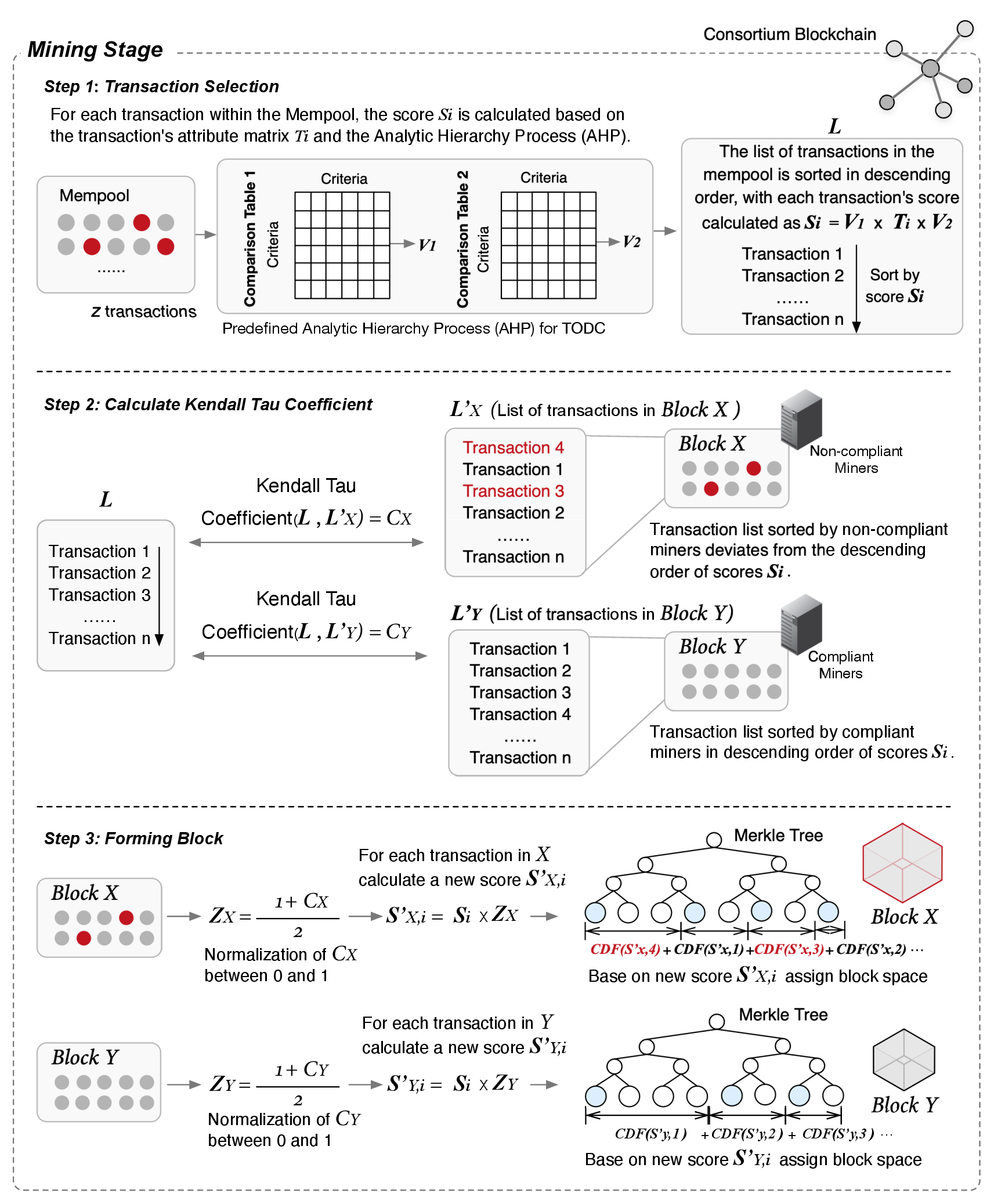}
\caption{Illustration of how Distributed Transaction Sequencing Strategy (DTSS) works in the mining stages}
    \label{DTSS2}
\end{figure*}

\item[Step2] \textbf{\textsl{Calculate Kendall Tau Coeffcient}}

To enhance the control over transaction sequencing within blocks, we employ the Kendall Tau Coefficient. This measure enables a comparison between the original transaction order in mempool and the miner's chosen sequence within a block, thereby guaranteeing alignment with necessary dependencies and the preservation of transaction effectiveness. For a more comprehensive understanding of the implementation mechanism, we detail the definitions of the variables utilized in Table \ref{table:variables}.

\begin{table}[H]
\footnotesize
\centering
\caption{Variables and their descriptions}
\resizebox{0.92\hsize}{!}{
\begin{tabular}{c|p{7.6cm}<{\centering}}
\hline\hline
        \specialrule{0.00em}{3pt}{1pt} 
\textbf{} & \textbf{Description} \\
    \specialrule{0.00em}{3pt}{1pt}  
    \hline
    \specialrule{0.00em}{3pt}{1pt} 
    $S_i$ & Original score for each transaction in the mempool \\
    \specialrule{0.00em}{3pt}{1pt}
    \hline
    \specialrule{0.00em}{3pt}{1pt} 
    $L$ & List of transactions in mempool sorted by descending order of scores $S_i$ \\
    \specialrule{0.00em}{3pt}{1pt} 
    \hline
    \specialrule{0.00em}{3pt}{1pt} 
    $L_X^{'}$ & Transaction list sorted by non-compliant miners deviates from the descending order of scores $S_i$\\
    \specialrule{0.00em}{3pt}{1pt} 
    \hline
    \specialrule{0.00em}{3pt}{1pt} 
    $C_X$ & Kendall tau coefficient for block $X$, within range -1 to 1 \\
    \specialrule{0.00em}{3pt}{1pt} 
    \hline
    \specialrule{0.00em}{3pt}{1pt} 
    $Z_X$ & Normalization of $C_X$ between 0 and 1, defined as $Z_X = \frac{(1+C_X)}{2}$ \\
    \specialrule{0.00em}{3pt}{1pt} 
    \hline
    \specialrule{0.00em}{3pt}{1pt} 
    $S'_{X,i}$ & New score for each transaction in block $X$, defined as $S'_{X,i} = {S_i}\times {Z_X}$ \\
    \specialrule{0.00em}{3pt}{1pt} 
    \hline
    \specialrule{0.00em}{3pt}{1pt} 
    $L_Y^{'}$ & Transaction list sorted by compliant miners in descending order of scores $S_i$\\
    \specialrule{0.00em}{3pt}{1pt} 
    \hline
    \specialrule{0.00em}{3pt}{1pt} 
    $C_Y$ & Kendall tau coefficient for block $Y$, expected value is 1 \\
    \specialrule{0.00em}{3pt}{1pt} 
    \hline
    \specialrule{0.00em}{3pt}{1pt} 
    $Z_Y$ & Normalization of $C_Y$ between 0 and 1, defined as $Z_Y = \frac{(1+C_Y)}{2}$ \\
    \specialrule{0.00em}{3pt}{1pt} 
    \hline
    \specialrule{0.00em}{3pt}{1pt} 
    $S'_{Y,i}$ & New score for each transaction in block $Y$, defined as $S'_{Y,i} = {S_i}\times {Z_Y}$ \\
    \specialrule{0.00em}{3pt}{1pt} 
    \hline
    \specialrule{0.00em}{3pt}{1pt} 
\end{tabular}
}
\label{table:variables}
\end{table}

Within the framework of DTSS, the initial step involves computing an individual score $S_i$ for each transaction in the mempool. Subsequently, a generalized list $L$ is generated, arranging the transactions in descending order based on their respective scores $S_i$. As miners proceed with the task of selecting and incorporating transactions into the block, they maintain the flexibility to adopt diverse rules governing the selection process from the mempool. It is important to recognize that the order in which miners choose transactions can take two distinct forms. The first form entails adhering to DTSS's guidelines and selecting transactions in accordance with the original list $L$, denoted as $L_Y'$. Conversely, miners may deviate from the original list and select transactions in a manner that diverges from DTSS, resulting in an ordered list denoted as $L_X'$.

Considering the scenario where $L_X^{'}$ does not obey the descending order of score $S_i$, a non-compliant miner incorporates these transactions into a block, denoted by $X$, without considering the original scores in $L$. The Kendall tau coefficient $C_X$ is then calculated based on the order of transactions in $L_X^{'}$ and $L$. This coefficient is a measure of rank correlation quantifying the degree of similarity between two rankings. This value of $C_X$ is then normalized to a value between 0 and 1 using the formula $Z_X = \frac{(1+C_X)}{2}$. A new score $S'_{X,i}$ for each transaction in block $X$ is then calculated using the formula $S'_{X,i} = S_i \times Z_X$. Due to the divergence between $L$ and $L_X^{'}$, it can be anticipated that the normalized coefficient $Z_X$ will fall below 1. This, in turn, yields a $S'_{X,i}$ that is lower than the original score $S_i$ for each transaction incorporated within block $X$.

On the other hand, another compliant miner strictly orders transactions from highest to lowest score to form $L_Y^{'}$. These transactions are then incorporated into a block, denoted by $Y$. Given the transactions are strictly ordered, the Kendall tau coefficient $C_Y$ for this block is computed to be 1. The value of $C_Y$ is normalized using the formula $Z_Y = \frac{(1+C_Y)}{2}$, resulting in $Z_Y$ also being 1. Finally, a new score $S'_{Y,i}$ for each transaction in block $Y$ is calculated using the formula $S'_{Y,i} = S_i \times Z_Y$. Considering the equivalence between $L$ and $L_Y^{'}$, it follows that $Z_Y$ equals 1. Consequently, the new score $S'_{Y,i}$ will correspond exactly to the original score $S_i$ for each transaction within block $Y$.

The Kendall tau coefficient, denoted as $\tau$, is computed as \refeqs{KendallTau}, where $n$ is the total number of pairs. In this formula, a pair is said to be concordant if the ranks for both elements agree in their order (for example, the higher-ranked element in the first ranking is also higher in the second ranking). Conversely, a pair is discordant if the ranks disagree in their order. The denominator, $\frac{1}{2}n(n-1)$, is the total number of pairs. The resulting coefficient ranges from -1, indicating that the two rankings are perfectly inverted, to +1, indicating that the two rankings are identical. A coefficient of 0 indicates no correlation between the rankings.

\begin{normalsize}
    \begin{equation}
        \displaystyle
        \resizebox{0.9\hsize}{!}{
                $\tau = \frac{(\text{number of concordant pairs}) - (\text{number of discordant pairs})}{\frac{1}{2}n(n-1)} $}
        \label{KendallTau}
    \end{equation}
\end{normalsize}

\item[Step3] \textbf{\textsl{Forming Block}}

In \reffig{DTSS2} \textsl{Mining Stage} \textsl{Step 1}, compliant miners select transactions to be incorporated in the block based on the order of scores, from lowest to highest. \textcolor{black}{Non-compliant miners, in pursuit of additional mining rewards through successful BEV attacks, may disregard the established mechanisms designed to select and rank transactions based on their scores.} The Cumulative Distribution Function (CDF) is an essential concept in probability theory and statistics, describing the likelihood that a random variable $X$ will take a value less than or equal to a given value $x$. As outlined in \refeqs{con:cdf} \textsl{Mining Stage} \textsl{Step 2}, DTSS employs the CDF of the log-normal distribution, denoted as $F(x)$, to calculate the leaf space corresponding to the score of the transaction selected by the miner. In \refeqs{con:cdf}, \textsl{Scale} ($\mu$) pertains to the mean of the naturally distributed logarithm, while \textsl{Shape} ($\sigma$) refers to the standard deviation of the naturally distributed logarithm. 

\begin{equation}
\displaystyle
\resizebox{.6\hsize}{!}{
$F(x) = Pr(X\leq x) = \frac{1}{2} \, + \, \frac{1}{2} \text{erf} \bigg[\frac{ln(x)-\mu}{\sigma\sqrt{2}} \bigg]$ \label{con:cdf}
}
\end{equation}

Leaf space reflects the number of Merkle tree leaf nodes that the transaction should occupy within the block. A transaction with a higher score $S_i$ will be allocated a greater leaf space, thereby occupying more nodes within the Merkle tree. Any additional leaf nodes that are reserved for this transaction but are not required due to the transaction's inherent size will remain empty. As a result, a block that incorporates a higher quantity of lower-scoring transactions, inclusive of those related to BEV attack transactions, accommodates a larger number of transactions. This leads to a larger block size, exemplified by $Block X'$ in \reffig{DTSS2}. In contrast, a block that incorporates higher-scoring transactions accommodates fewer transactions, resulting in a smaller block size, as demonstrated by $Block X$ in \reffig{DTSS2}.

\end{description}

\subsection{Propagation Stage}

\begin{figure*}
\centering
\includegraphics[width=0.7\linewidth]{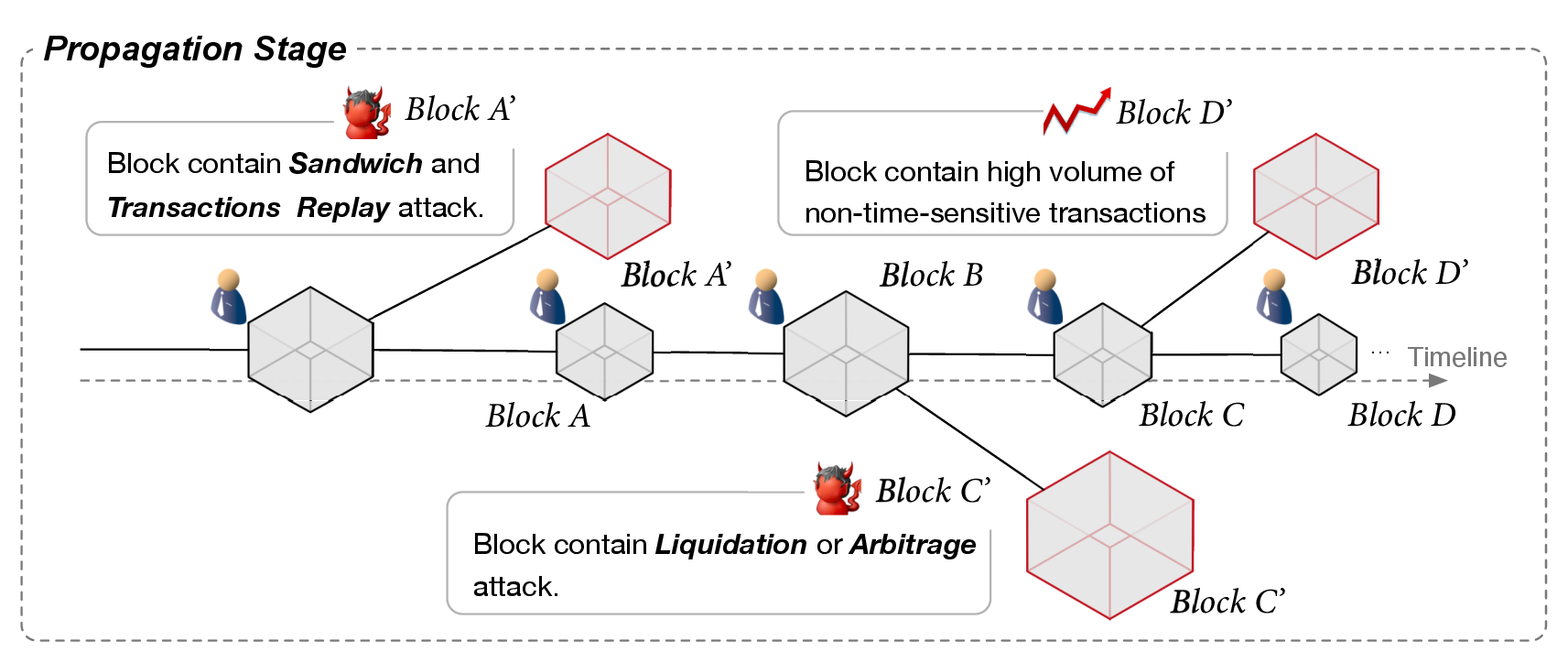}
\caption{Illustration of how Distributed Transaction Sequencing Strategy (DTSS) works in the propagation stages}
    \label{DTSS3}
\end{figure*}

\textcolor{black}{As depicted in the \textsl{Propagation Stage} of \reffig{DTSS3}, blocks with BEV or non-time-sensitive TID transactions and those without are mined concurrently in different stages. Smaller blocks, often containing fewer transactions, reach Blockchain network consensus faster, thus joining the longest chain. However, blocks containing aggressive strategies such as Sandwich attacks, Liquidations, arbitrage attacks, and Transaction Replay attacks are limited by DTSS, resulting in non-competitive block sizes. This stems from the transaction selection and ranking process during Mining Stage, Step 1 and Step 2 in \reffig{DTSS2}, where compliant miners follow established rules while non-compliant miners might bypass them for extra rewards. This process, facilitated by the Cumulative Distribution Function (CDF) (\refeqs{con:cdf}) in Mining Stage, Step 2, leads to block space variance. Consequently, blocks from non-compliant miners are often forked off, as compliant miners' blocks form the longest chain. This strategic transaction ordering effectively mitigates BEV attacks and prioritizes time-sensitive TID transactions.}

\noindent

The aforementioned pseudocode (\refargo{DTSSAlgo}) presents a detailed procedural representation of DTSS. 

\begin{breakablealgorithm}
\begin{footnotesize}
\floatname{algorithm}{Pseudo-code}
\caption{DTSS}
\label{DTSSAlgo}
\begin{algorithmic}[1]
\State \textbf{Input:} Mempool, AHP comparison tables
\State \textbf{Output:} Mined block $B$
\Procedure{DTSS}{}
    \State Initialize two AHP comparison tables and compute priority vectors $V1$ and $V2$
    \State Initialize empty lists $L$, $L_{X'}$, and $L_{Y'}$
    \State \texttt{\textbackslash{}\textbackslash{} Mining Stage - Step 1: Transaction Selection}
    \For{each transaction in mempool}
        \State Compute the score for each transaction, $S_i$, using AHP tables and priority vectors $V1$ and $V2$
        \State Add each transaction and its corresponding score to list $L$
    \EndFor
    \State Sort list $L$ in descending order of transaction scores $S_i$
    \For{each miner}
        \If{miner does not adhere to DTSS}
            \State \texttt{\textbackslash{}\textbackslash{} Mining Stage - Step 2: Calculate Kendall Tau Coefficient}
            \State Select transactions from mempool in an order diverges from list $L$ to form $L_{X'}$
            \State Compute the Kendall tau coefficient $C_X$ based on $L_{X'}$ and $L$
            \State Normalize Kendall tau coefficient to a value between 0 and 1: $Z = (1 + C_X) / 2$
            \State Compute the new score for each transaction in the block: $S'_{X,i} = S_i * Z$
            \State \texttt{\textbackslash{}\textbackslash{} Mining Stage - Step 3: Forming Block}
            \For{each transaction $T_i$ in $L_{X'}$}
            \State Calculate leaf space $Leaf_i$ for $T_i$ using CDF of log-normal distribution
            \State $Leaf_i = \frac{1}{2} + \frac{1}{2} \text{erf} \left[\frac{\ln(S'_{X,i})-\mu}{\sigma\sqrt{2}} \right]$
                \If{$\sum Leaf_i>2100$}
                    \State break
                \Else
                    \State Allocate $Leaf_i$ in $X$ for $T_i$
                \EndIf
            \EndFor
        \Else
            \State \texttt{\textbackslash{}\textbackslash{} Mining Stage - Step 2: Calculate Kendall Tau Coefficient}
            \State Select transactions from list $L$ in descending order of score $S_i$ to form $L_{Y'}$
            \State Compute the Kendall tau coefficient $C_Y$ based on $L_{Y'}$ and $L$
            \State Normalize Kendall tau coefficient to a value between 0 and 1: $Z = (1 + C_Y) / 2$
            \State Compute the new score for each transaction in the block: $S'_{Y,i} = S_i * Z$
            \State \texttt{\textbackslash{}\textbackslash{} Mining Stage - Step 3: Forming Block}
            \For{each transaction $T_i$ in $L_{Y'}$}
            \State Calculate leaf space $Leaf_i$ for $T_i$ using CDF of log-normal distribution
            \State $Leaf_i = \frac{1}{2} + \frac{1}{2} \text{erf} \left[\frac{\ln(S'_{Y,i})-\mu}{\sigma\sqrt{2}} \right]$
                \If{$\sum Leaf_i>2100$}
                    \State break
                \Else
                    \State Allocate $Leaf_i$ in $Y$ for $T_i$
                \EndIf  
            \EndFor
        \EndIf
    \EndFor
    \State \texttt{\textbackslash{}\textbackslash{} Propagate Stage}
    \For{each block $B$ in [$X$, $Y$ ...]}
        \If{$B$ reaches consensus}
            \State Add $B$ to the longest chain
        \Else
            \State Fork $B$ off the chain
        \EndIf
    \EndFor
\EndProcedure
\end{algorithmic}
\end{footnotesize}
\end{breakablealgorithm}

In the subsequent section, we describe the assembly of a comparison table employing the Analytic Hierarchy Process (AHP). This procedure forms the central component of DTSS. It is specifically designed to enable miners to compliantly select transactions for incorporation in a block, taking into account a variety of transaction attributes.

\subsection{Analytic Hierarchy Process (AHP)}

\begin{itemize}

\item \textbf{\textsl{Step 1:}} To address the issues associated with BEV and TID, we establish four criteria for transaction evaluation: \textbf{Transaction Type}, includes categories such as Payment, Security Trading, Transfer, Instruction, and Statement. \textbf{Transaction Amount}, refers to the monetary value of the transaction. \textbf{Fee Percentage}, represents the proportion of the transaction amount that is allocated as a fee. \textbf{Initiation Time}, refers to the timestamp when the transaction was initiated.

\item \textbf{\textsl{Step 2:}} After identifying the transaction criteria, we next construct comparison tables based on these criteria, to capture the relative importance of each criteria with respect to all other criteria to prioritize and weigh transactions. The entries in the table represent the degree to which one criteria is judged to be more important than another. Given that Transaction Type is a discrete variable, whereas Transaction Fee, Fee Percentage, and Initiate Time are continuous variables, we introduce two pairwise comparison tables associated with the AHP algorithm within DTSS. 

Our initial approach employs a pairwise comparison table, presented in Table \ref{cmpt1.1}, which introduces importance level variables \textbf{A1-A10}. These variables are explicitly crafted to consider the relative priorities of various Transaction Types, encompassing Payment, Security Trading, Transfer, Instruction, and Statement. This strategic configuration aids in addressing the complexities inherently associated with TID.

In the financial sector, transaction types are prioritized based on factors such as urgency, value, and risk. Security Trading transactions are prioritized due to the dynamic nature of financial markets where delayed execution can result in significant financial losses. Instruction transactions, setting the groundwork for subsequent financial activities, are prioritized over Payment transactions, which, despite being crucial, are usually routine and have less immediate impact. Payment transactions, representing the settlement of debts or completion of transactions, take precedence over Transfer transactions, which are often internal movements of funds without an immediate obligation. Lastly, Transfer transactions, which involve the actual movement of funds, are prioritized over Statement transactions, which are generally informational and do not directly influence the financial position. This hierarchy reflects the critical need for timeliness and accuracy in the financial sector.

\textcolor{black}{Within DTSS, we have implemented transaction type restrictions, setting a precedence hierarchy: Security Trading supersedes Instruction, which precedes Payment. Payment then holds priority over Transfer, followed by Statement.}


\begin{table}[H]
        \footnotesize
        \centering
        \caption{\textcolor{black}{Pairwise Comparison Table Formed by $A_1$ to $A_{10}$ for Criteria 1 (Transaction Type)}}
        \resizebox{0.98\hsize}{!}{
        \label{cmpt1.1}
        \begin{tabular}{c|c|c|c|c|c}
        \hline\hline
        \specialrule{0.00em}{3pt}{1pt} 
         & Payment & Security Trading & Transfer & Instruction & Statement\\
        \specialrule{0.00em}{3pt}{1pt} 
        \hline
        \specialrule{0.00em}{3pt}{1pt} 
        Payment & 1 & $\frac{1}{A1}$ & $\frac{1}{A2}$ & $\frac{1}{A4}$ & $\frac{1}{A7}$\\
        \specialrule{0.00em}{3pt}{1pt} 
        \hline
        \specialrule{0.00em}{3pt}{1pt} 
        Security Trading & $A1$ & 1 & $\frac{1}{A3}$ & $\frac{1}{A5}$ & $\frac{1}{A8}$ \\
        \specialrule{0.00em}{3pt}{1pt} 
        \hline
        \specialrule{0.00em}{3pt}{1pt} 
        Transfer & $A2$ & $A3$ & 1 & $\frac{1}{A6}$ & $\frac{1}{A9}$\\
        \specialrule{0.00em}{3pt}{1pt} 
        \hline
        \specialrule{0.00em}{3pt}{1pt} 
        Instruction & $A4$ & $A5$ & $A6$ & 1 & $\frac{1}{A{10}}$\\
        \specialrule{0.00em}{3pt}{1pt} 
        \hline
        \specialrule{0.00em}{3pt}{1pt} 
        Statement & $A7$ & $A8$ & $A9$ & $A{10}$ & 1\\
        \specialrule{0.00em}{3pt}{1pt} 
        \hline
        \end{tabular}}  
        \begin{tablenotes}
        \footnotesize
        \item \textcolor{black}{Each of the following represents the importance level of one transaction type over another: \textbf{\textsl{A1}} Security Trading over Payment, \textbf{\textsl{A2}} Transfer over Payment, \textbf{\textsl{A3}} Transfer over Security Trading, \textbf{\textsl{A4}} Instruction over Payment, \textbf{\textsl{A5}} Instruction over Security Trading, \textbf{\textsl{A6}} Instruction over Transfer. Similarly, \textbf{\textsl{A7}} Transfer over Payment again, \textbf{\textsl{A8}} Statement over Security Trading, \textbf{\textsl{A9}} Statement over Transfer, \textbf{\textsl{A10}} Statement over Security Trading.}
        \end{tablenotes}
    \end{table}

Subsequent to the first table, we present a second pairwise comparison table, delineated in Table \ref{cmpt2.1}. This table introduces importance level variables \textbf{A11-A13}. These variables are particularly designed to contemplate the varying priorities of different transaction attributes, including Transaction Fee, Fee Percentage, and Initiation Time. This strategic configuration aids in mitigating the potential risks of BEV attacks.


\begin{table}[htp]
        \footnotesize
        \centering
        \caption{\textcolor{black}{Pairwise Comparison Table Formed by $A{11}$ to $A{13}$ for Criteria 2 (Transaction Amount), Criteria3 (Initiation Time) and Criteria 4 (Fee Percentage)}}
        \resizebox{0.98\hsize}{!}{
        \begin{tabular}{c|c|c|c}
        \hline\hline
        \specialrule{0.00em}{3pt}{1pt} 
        & Transaction Amount & Initiation Time & Fee Percentage\\
        \specialrule{0.00em}{3pt}{1pt} 
        \hline
        \specialrule{0.00em}{3pt}{1pt} 
        Transaction Amount & 1 & $\frac{1}{A{11}}$ & $\frac{1}{A{12}}$ \\
        \specialrule{0.00em}{3pt}{1pt} 
        \hline
        \specialrule{0.00em}{3pt}{1pt} 
        Initiation Time & $A{11}$ & 1 & $\frac{1}{A{13}}$\\
        \specialrule{0.00em}{3pt}{1pt} 
        \hline
        \specialrule{0.00em}{3pt}{1pt} 
        Fee Percentage & $A{12}$ & $A{13}$ & 1\\
        \specialrule{0.00em}{3pt}{1pt} 
        \hline
        \specialrule{0.00em}{3pt}{1pt} 
        \end{tabular}}
        \begin{tablenotes}
        \footnotesize
        \item \textcolor{black}{Each of the following represents the importance level of one transaction attribute over another: \textbf{\textsl{A11}}  Initiation Time over Transaction Amount, \textbf{\textsl{A12}} Fee Percentage over Transaction Amount, \textbf{\textsl{A13}} Fee Percentage over Initiation Time.}
        \end{tablenotes}
        \label{cmpt2.1}
\end{table}

\item \textbf{\textsl{Step 3:}} Upon constructing the two pairwise comparison tables, we calculate the priority vector for different transactions. The priority vector signifies the relative importance of each criteria and can be utilized to rank the weight of transactions for final decision-making. To derive the priority vector, we normalize the data in Table \ref{cmpt1.1} and Table \ref{cmpt2.1} by dividing each element by the sum of its corresponding column. This process results in the relative importance weight of the criteria, which are the sum of the entries within the same row. This method offers a systematic and quantifiable approach to discern the relative significance of each transaction based on predefined criteria.


\item \textbf{\textsl{Step 4:}} To derive a quantitative assessment of each transaction, a score is calculated by multiplying the normalized weight of each criterion with corresponding transaction attributes, and subsequently aggregating these across all criteria.

\item \textbf{\textsl{Step 5:}} With the priority vector calculated, miners can proceed to select transactions based on the weighted sum of scores in accordance with the consensus protocol. This method provides a structured and quantitative means to prioritize transactions, ensuring that those with higher weights are selected for incorporation in the block first.

\end{itemize}

The Analytic Hierarchy Process (AHP) has been chosen in this paper due to its simplicity and the interpretability of its results. The AHP uses a structured approach to break down a complex decision problem into a hierarchy of more easily comprehensible sub-problems, each of which can be analyzed independently. The resulting prioritized ranking of decision criteria is easy to understand and interpret, making AHP a particularly suitable choice for applications in the financial sector where interpretability and transparency are crucial. \textcolor[RGB]{0,0,0}{The DTSS method, with its AHP-based weight assignment and sorting mechanism, introduces additional computational overhead compared to simpler methods like FIFO. The eigenvalue computation in AHP can be more time-consuming $(O(n^2))$, as the primary computational burden lies in constructing and normalizing the pairwise comparison matrix, which involves $n^2$ elements.} Indeed, there exists a plethora of Multiple Criteria Decision Making (MCDM) techniques, each offering a unique methodology for evaluating and selecting among different alternatives based on multiple criteria. Nonetheless, these techniques vary significantly in terms of their complexity, interpretability, and suitability for specific applications. Hence, the selection of a particular technique must be tailored to the specific requirements and context of the application at hand.

For instance, the Analytic Network Process (ANP) \cite{saaty2013analytic} is an extension of the AHP that considers interdependencies among decision levels and alternatives, making it more complex than AHP. Techniques like the Technique for Order of Preference by Similarity to Ideal Solution (TOPSIS) \cite{tzeng2011multiple} and Elimination and Choice Expressing Reality (ELECTRE) \cite{figueira2013overview} provide comprehensive ranking and decision-making frameworks, but they may be difficult to interpret and apply in certain contexts. On the other hand, methods such as Simple Additive Weighting (SAW) \cite{afshari2010simple} and Multi-Attribute Utility Theory (MAUT) \cite{von1975multi} are relatively simple and straightforward but may not capture the sophisticated nuances of certain decision problems. Techniques such as VlseKriterijumska Optimizacija I Kompromisno Resenje (VIKOR) \cite{halim2022vise}, Data Envelopment Analysis (DEA) \cite{bowlin1998measuring}, and Goal Programming (GP) \cite{tamiz1998goal} have specific applications in areas like optimization and production, which may not be directly applicable in all decision-making scenarios.

Following a comprehensive overview of DTSS, one crucial matter that still requires attention is determining the value of the importance level (A1-A13) in the AHP comparison table. In the following section, our objective is to identify the optimal setting for the importance level value using algorithm optimization techniques. This endeavor aims to enhance the effectiveness of addressing BEV attacks and TID challenges, providing a more efficient resolution to these issues.

\section{Experiments}

In \textcolor{black}{this section, the experiment optimize the importance level setting in the AHP comparison table under DTSS using optimization algorithms and SimBlock \cite{8751431}. It aims to compare DTSS with other BEV solutions for analysis.} To better simulate real-world DeFi transaction scenarios, we describe the methodology used to generate the experimental data. Subsequently, we delve into the iterative optimization of the importance level setting in AHP comparison table using optimization algorithms, in conjunction with SimBlock \cite{8751431}. The objective is to attain the optimal setting under DTSS. Lastly, we compare DTSS against other well-established solutions for BEV, enabling a comparative analysis.

\subsection{Experiment Settings}

\textcolor{black}{Our experimental platform was custom-tailored utilizing the resources provided by Huawei Cloud Stack (HCS), with the support of the Super Intelligent Computing Center (SICC) at the University of Macau. We conducted our experiments on a Ubuntu Linux v18.04 virtual machine equipped with 48 vCPUs at 2.50GHz, 192GB of RAM, and 1TB of ROM. To evaluate the effectiveness of DTSS in controlling the order in which transactions are incorporated in the blocks, we utilized SimBlock \cite{8751431}, a tool widely recognized for its configurability in terms of hash power distribution, network latencies, incentive mechanisms, and more. In the default setting of SimBlock, blocks are generated based on a probability model that assumes Proof-of-Work. These blocks are then propagated throughout the simulated Blockchain network. This integrated and meticulously configured setup ensured accurate and meaningful experimental results.}

The Bitcoin protocol imposes specific constraints on block size, capping it at 1 MB. A block should include all valid transactions if the available transactions do not fully occupy a 1 MB block. Typically, transactions contain around 500 bytes of data. For our experiments, we adhered to Bitcoin's average number of transactions per block over a year, which is approximately 2,100 transactions. This approach ensured our experimental setup accurately reflected real-world Bitcoin transaction dynamics.

Within the proposed simulation framework, we have examined multiple optimization algorithms to generate a collection of candidate solutions representing various settings for DTSS. Specifically, we have evaluated three optimization algorithms: Differential Evolution (DE) \cite{5601760}, Particle Swarm Optimization (PSO) \cite{eberhart1995new}, and Gradient-based Optimizers (GBO) \cite{AHMADIANFAR2020131}. These algorithms have been tested for their efficacy in our simulation framework. The optimization algorithms employ a population size of 50 and a maximum generation limit of 100. For the DE algorithm, we have set the crossover constant to 0.9, which is deemed suitable for attributes with dependencies. Additionally, the GBO algorithm's local escaping probability is set to 0.5, as suggested in \cite{AHMADIANFAR2020131}.

\begin{table}[htp]
\footnotesize
\caption{\textcolor{black}{Parameters for Optimization Algorithms}}
\label{opt_param}
\centering
\resizebox{0.98\hsize}{!}{
\begin{tabular}{>{\color[RGB]{0,0,0}}p{1.2cm}<{\centering} |  >{\color[RGB]{0,0,0}}p{2.2cm}<{\centering} | >{\color[RGB]{0,0,0}}p{2.6cm}<{\centering} | >{\color[RGB]{0,0,0}}p{0.8cm}<{\centering}}
\hline\hline
\textbf{Algorithm} & \textbf{Hyper-parameter} & \textbf{description} & \textbf{value} \\ 
\specialrule{0.00em}{0pt}{0pt}
\hline
\multirow{3}{*}{DE}  & n\_pop          & population                 & 50    \\ 
\specialrule{0em}{0pt}{0pt}\cline{2-4}
                     & max\_gen        & maximum generation         & 100   \\ \specialrule{0em}{0pt}{0pt}\cline{2-4}
                     & CR              & crossover constant         & 0.9   \\ 
                     \specialrule{0em}{0pt}{0pt}\hline
\multirow{2}{*}{
GA}  & n\_pop          & population                 & 50    \\ 
                     \specialrule{0em}{0pt}{0pt}\cline{2-4} 
                     & max\_gen        & maximum generation         & 100   \\ 
                     \specialrule{0em}{0pt}{0pt}\hline
\multirow{5}{*}{PSO} & n\_pop          & population                 & 50    \\ 
                     \specialrule{0.00em}{0pt}{0pt}\cline{2-4}
                     & max\_gen        & maximum generation         & 100   \\ 
                     \specialrule{0.00em}{0pt}{0pt}\cline{2-4}
                     & w               & inertia weight             & 0.73  \\ 
                     \specialrule{0.00em}{0pt}{0pt}\cline{2-4}
                     & c1              & acceleration coefficient   & 1.50  \\ 
                     \specialrule{0.00em}{0pt}{0pt}\cline{2-4}
                     & c2              & acceleration coefficient   & 1.50  \\ 
                     \specialrule{0.00em}{0pt}{0pt}\hline
\end{tabular}
}
\end{table}

\subsection{Experiment Data}

In our study, we sought to evaluate the robustness of our proposed algorithm against potential real-world disruptions. We utilized historical market data from Bitcoin, spanning from December 2019 to September 2020. This dataset comprised 400,000 transactions, with minute-to-minute updates on OHLC (Open, High, Low, Close), transaction amount in BTC, and weighted Bitcoin price. To validate the efficacy of DTSS in addressing TID issue, we initially classify the normal transaction data into distinct categories such as Payment, Security Trading, Transfer, Instruction, and Statement. Subsequently, we integrate abnormal transactions into the normal transaction dataset. This integration aims to simulate a diverse range of BEV attacks, with specific emphasis on Sandwich Attacks, Replay Attacks, and Clogging. 

The BEV attack transactions were systematically introduced based on a predefined set of rules. We implemented a threshold to identify the top 1\% of transactions as potential victim transactions. For each percentile of identified victim transactions, we inserted two transactions for Sandwich attacks and one transaction for replay attacks. The transaction amount for these BEV attack transactions was randomly assigned, with transaction fees typically set higher than those of normal transactions. Furthermore, careful control was exercised over the timing of these transactions to ensure their initiation after the victim's transaction had taken place. 

\begin{figure}[H]
    \makeatletter
    \def\@captype{figure} 
    \makeatother
    \centering
    \includegraphics[width=0.4 \textwidth]{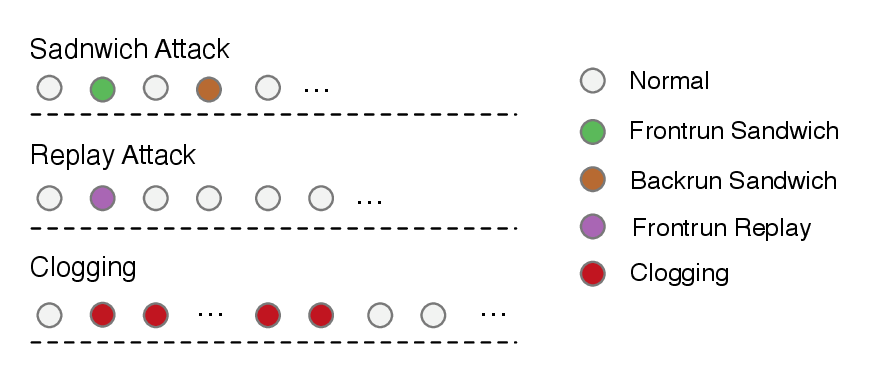}
    \caption{Illustration of how to insert adversary transaction to simulate Sandwich attack, replay attack, and clogging}
    \label{dataEnhancement}
\end{figure}

The diagram in Figure \ref{dataEnhancement} illustrates our methodology for simulating Sandwich Attack, Replay Attack, and Clogging by inserting adversary transactions. In the case of a \textbf{Sandwich Attack}, we emulate this scenario by configuring a pair of adversary transactions with higher fees to surround a victim transaction. These represent the front-run and back-run transactions. For a \textbf{Replay Attack}, a decoy transaction is inserted to replicate the victim transaction's execution logic, coupled with a pair of adversary transactions to extract profits. In simulating \textbf{Clogging}, adversary transactions are introduced to congest the Blockchain, thereby inhibiting users and bots from initiating transactions.

\vspace{-0.6cm}
\subsection{\textcolor{black}{Parameters} in DTSS}

In DTSS, \textcolor{black}{tables (Table \ref{cmpt1.1} \ref{cmpt2.1}) are} created to evaluate each criterion's relative importance, aiding in the weighting and prioritization of transactions. Among all entries, there are 13 distinct importance level settings (\textbf{A1-A13}) that contribute to the final score of each transaction. The block space each transaction occupies is determined using the Cumulative Distribution Function (CDF) algorithm, with the optimization of two parameters Scale and Shape ensuring optimal block space allocation, thereby enhancing the efficacy of DTSS.

\begin{figure}[H]
\makeatletter
\def\@captype{figure}
\makeatother
\centering
\includegraphics[width=0.34\textwidth]{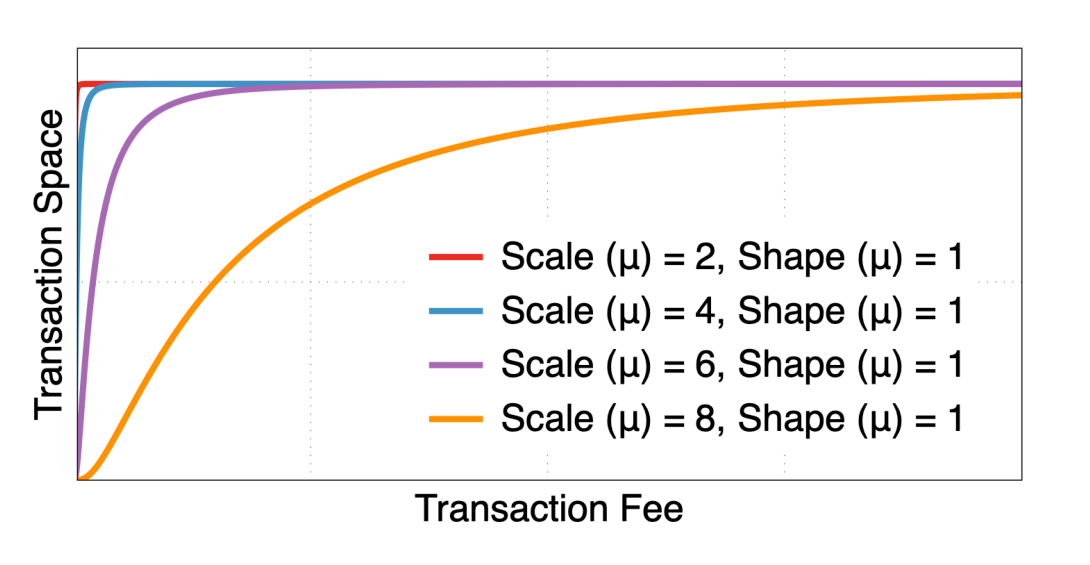}
\caption{Storage space allocation based on score with different Scale ($\mu$) parameter under TDOC}
\label{CDFParam1}
\end{figure}

\textbf{\textsl{Scale:}} The \textsl{Scale} ($\mu$) in \refeqs{con:cdf} is the mean of the normally distributed natural logarithm. The value of the \textsl{Scale} parameter determines the statistical dispersion or ``scale'' of the probability distribution. As shown in \reffig{CDFParam1}, if the \textsl{Scale} parameter is small, the probability distribution is more concentrated. The probability distribution is more dispersed if the \textsl{Scale} parameter is large.

\begin{figure}[H]
\makeatletter
\def\@captype{figure}
\makeatother
\centering
\includegraphics[width=0.34\textwidth]{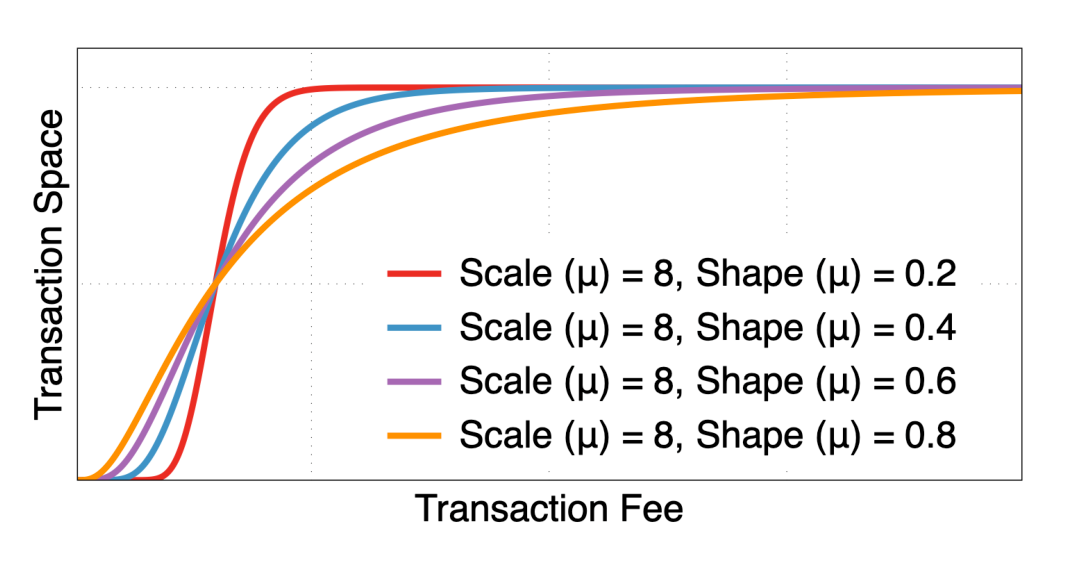}
\caption{Storage space allocation based on score with different Shape ($\sigma$) parameter under TDOC}
\label{CDFParam2}
\end{figure}

\textbf{\textsl{Shape:}} The \textsl{Shape} ($\sigma$) in  \refeqs{con:cdf} is the standard deviation of the normally distributed natural logarithm. As shown in \reffig{CDFParam2}, \textcolor{black}{when the shape parameter values are different, under the same transaction fee level, the probability distributions of the maximum space occupied by transactions are also different.}

\subsection{Benchmarks}
In this study, we introduce the Normalized Allocation Disparity Metric (NADM) as an experimental benchmark to evaluate and optimize the resource allocation in our Blockchain mechanism. The metric is calculated by \refeqs{NADM}
\begin{equation}
\label{NADM}
\displaystyle
\resizebox{.6\hsize}{!}{
    $NADM = \frac{Mean(Leaf_{normal})-Mean(Leaf_{adversary})}{Max(Leaf)}$
}
\end{equation}

\textcolor{black}{Where Mean($Leaf_{normal}$) is the average space allocated to normal transactions, Mean($Leaf_{adversary}$) is the average space allocated to adversary transactions and Max($Leaf$) is the maximum space a transaction can occupy.}

The rationale behind using this metric is to assess the impact of resource allocation on the incentives and penalties faced by adversary miners. A larger benchmark value indicates a greater disparity between normal and adversary transactions, ultimately resulting in larger block sizes for adversary blocks. This, in turn, increases the vulnerability of adversary blocks to forking and reduces their profitability. \textcolor{black}{Hence, a larger NADM value indicates superior performance.} By optimizing this metric, we aim to impose more significant penalties on adversary miners and enhance the security and stability of the Blockchain system. Through our experiments, we evaluate the performance of different parameter settings and optimization techniques in achieving a desirable level of allocation disparity and mitigating the influence of adversary transactions on the Blockchain's integrity.

\subsection{Optimization process}

\begin{figure}[H]
    \makeatletter
    \def\@captype{figure} 
    \makeatother
    \centering
    \includegraphics[width=0.5 \textwidth]{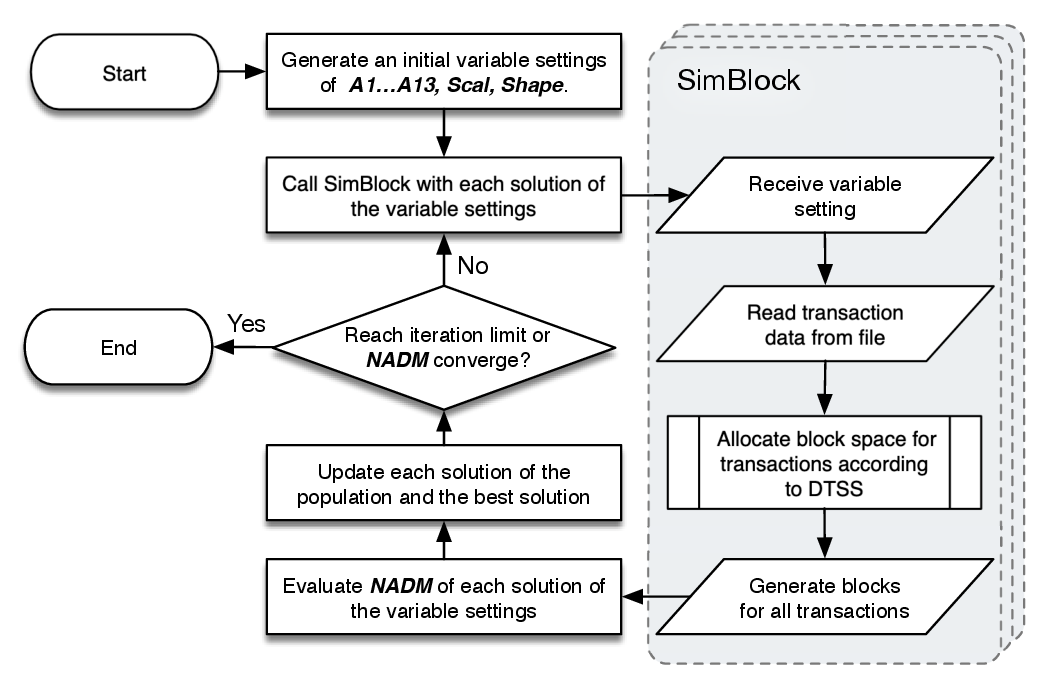}
    \caption{Schematic diagram of DTSS variable optimization experiment.}
    \label{DTSSProcess}
\end{figure}

As shown in \reffig{DTSSProcess}, the optimization experimental process begins with an initialization phase, where we generate an initial set of candidate solutions. Each solution represents a unique combination of variable settings for $A1$ through $A13$, along with the $Scale$ and $Shape$ parameters for DTSS. \textcolor{black}{During optimization, we uphold the A1-A10 parameter hierarchy, ensuring execution priority as: Security Trading $>$ Instruction $>$ Payment $>$ Transfer $>$ Statements.} Following initialization, we execute the optimization algorithm. Candidate solutions are passed as arguments to a simulator, which evaluates their performance. This simulation stage employs a Blockchain simulator to mimic the transaction incorporation behaviors of miners under varying DTSS variable settings.

After the simulation, we proceed to the evaluation stage. Here, the performance of candidate solutions is assessed by analyzing the simulation results. The optimizer gauges the Normalized Allocation Disparity Metric (NADM), which quantifies the difference between the average space allocated to normal and adversarial transactions. If a particle's NADM is superior to its previous value, the local best of that particle is updated. Concurrently, the global best is updated based on all particles' performance. The optimization process continues through iterative adjustments. New candidate solutions are generated, and the execution-simulation-evaluation cycle repeats until a predefined stopping criterion, such as a maximum number of iterations or a desired level of NADM convergence, is reached.

\subsection{Experimental results}

We conducted several experiments using different optimization algorithms and evaluated their effectiveness in mitigating adversarial transactions. Our experiments also included simulations of transaction executions with and without DTSS, allowing us to visually represent the significant reduction of adversarial transactions when DTSS is applied. The Independent Samples T-Test was used to statistically validate the impact of DTSS on transaction execution priority, revealing a significant influence across various transaction types. Subsequently, we also analyzed the effect of forking threats on DTSS, simulating the Mean Propagation Duration of blocks of different sizes within a blockchain network of 6000 nodes.

\subsubsection{Optimization results}

Table \ref{optimization_res} presents the outcomes of three distinct optimization algorithms, with each algorithm being evaluated through a separate experiment. The results demonstrate that Experiment 2, which employed the GA optimization algorithm, shows the most promising approach for mitigating adversarial transactions. This conclusion is supported by its higher NADM value of 0.931, which surpasses the values obtained in all other conducted experiments.

\begin{table*}
\footnotesize
\caption{\textcolor[RGB]{0,0,0}{Optimization Results with DE, PSO, and GBO}}
\label{optimization_res}
\centering
\resizebox{0.96\hsize}{!}{
\begin{threeparttable}
\begin{tabular}{>{\color[RGB]{0,0,0}}p{1.4cm}<{\centering}  >{\color[RGB]{0,0,0}}p{0.6cm}<{\centering}  >{\color[RGB]{0,0,0}}p{0.6cm}<{\centering}  >{\color[RGB]{0,0,0}}p{0.6cm}<{\centering}  >{\color[RGB]{0,0,0}}p{0.6cm}<{\centering}  >{\color[RGB]{0,0,0}}p{0.6cm}<{\centering}  >{\color[RGB]{0,0,0}}p{0.6cm}<{\centering}  >{\color[RGB]{0,0,0}}p{0.6cm}<{\centering}  >{\color[RGB]{0,0,0}}p{0.6cm}<{\centering}  >{\color[RGB]{0,0,0}}p{0.6cm}<{\centering}  >{\color[RGB]{0,0,0}}p{0.6cm}<{\centering}  >{\color[RGB]{0,0,0}}p{0.6cm}<{\centering}  >{\color[RGB]{0,0,0}}p{0.6cm}<{\centering}  >{\color[RGB]{0,0,0}}p{0.6cm}<{\centering}  >{\color[RGB]{0,0,0}}p{0.6cm}<{\centering}  >{\color[RGB]{0,0,0}}p{0.6cm}<{\centering} >{\color[RGB]{0,0,0}}p{1.2cm}<{\centering}}
\hline\hline
\textbf{Algorithm} & \textbf{A1} & \textbf{A2} & \textbf{A3} & \textbf{A4} & \textbf{A5} & \textbf{A6} & \textbf{A7} & \textbf{A8} & \textbf{A9} & \textbf{A10} & \textbf{A11} & \textbf{A12} & \textbf{A13} & \textbf{Scale} & \textbf{Shape}  & \textbf{NADM} \\ 
\specialrule{0em}{3pt}{1pt}\hline
DE &  0.83  & 0.63 & 0.94 & 0.51 & 0.76 & 0.15 & 0.82 & 0.12 & 0.93  & 0.66  & 0.41  & 10000  & 10000  & 5.42 & 0.3 & 0.926   \\ 
\specialrule{0em}{3pt}{1pt}\hline
\textbf{GA} &  \textbf{0.68}  & \textbf{0.84} & \textbf{0.72} & \textbf{0.69} & \textbf{0.73} & \textbf{0.33} & \textbf{0.98} & \textbf{0.74} & \textbf{0.96}  & \textbf{0.10}  & \textbf{0.90}  & \textbf{9952}  & \textbf{9644}  & \textbf{5.42} & \textbf{0.16} & \textbf{0.931}     \\  
\specialrule{0em}{3pt}{1pt}\hline
PSO &  0.73  & 0.77 & 0.97 & 0.53 & 0.74 & 0.17 & 0.92 & 0.33 & 0.94  & 0.27  & 0.51  & 9977  & 9958  & 5.94 & 0.13 & 0.912      \\ 
\specialrule{0em}{3pt}{1pt}
\toprule
\end{tabular}
\begin{tablenotes}
\footnotesize
\item A1-A10: Different importance level variable for Transaction Types, encompassing Payment, Security Trading, Transfer, Instruction, and Statement in Table \ref{cmpt1.1}; A11-A13: Different importance level variable for transaction attributes, including Transaction Fee, Fee Percentage, and Initiation Time in Table \ref{cmpt2.1}; 
\end{tablenotes}
\end{threeparttable}
}
\end{table*}

\begin{figure*}
    \makeatletter
    \def\@captype{figure} 
    \makeatother
    \centering
    \includegraphics[width=0.86 \textwidth]{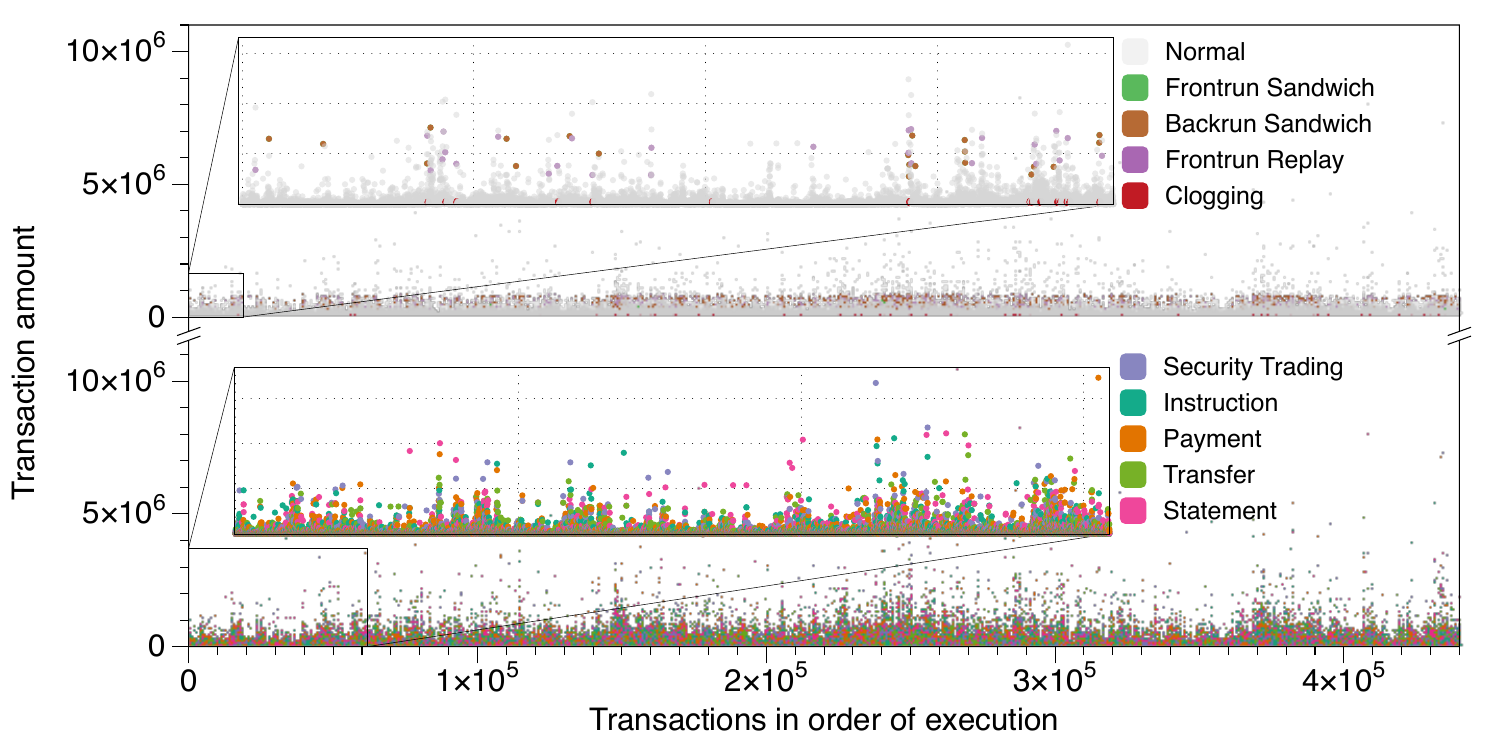}
    \caption{Illustration of the relationship between block size and the probability of multiple conflicting blocks being mined during block propagation (Probability of forking).}
    \label{TODCTransactions1}
\end{figure*}

To provide a more intuitive illustration of the impact of DTSS with optimized parameter settings, we first conduct a simulation of transactions without the application of DTSS. The simulation outcomes, as represented in \reffig{TODCTransactions1}, display the incorporation of normal and adversarial transactions within the block, based on the simulation dataset. The upper section of \reffig{TODCTransactions1} illustrates the distribution of adversarial transactions, whereas the lower section portrays the dispersion of various transaction types, such as Security Trading, Instruction, Payment, Transfer, and Statement.

\begin{figure*}
    \makeatletter
    \def\@captype{figure} 
    \makeatother
    \centering
    \includegraphics[width=0.86 \textwidth]{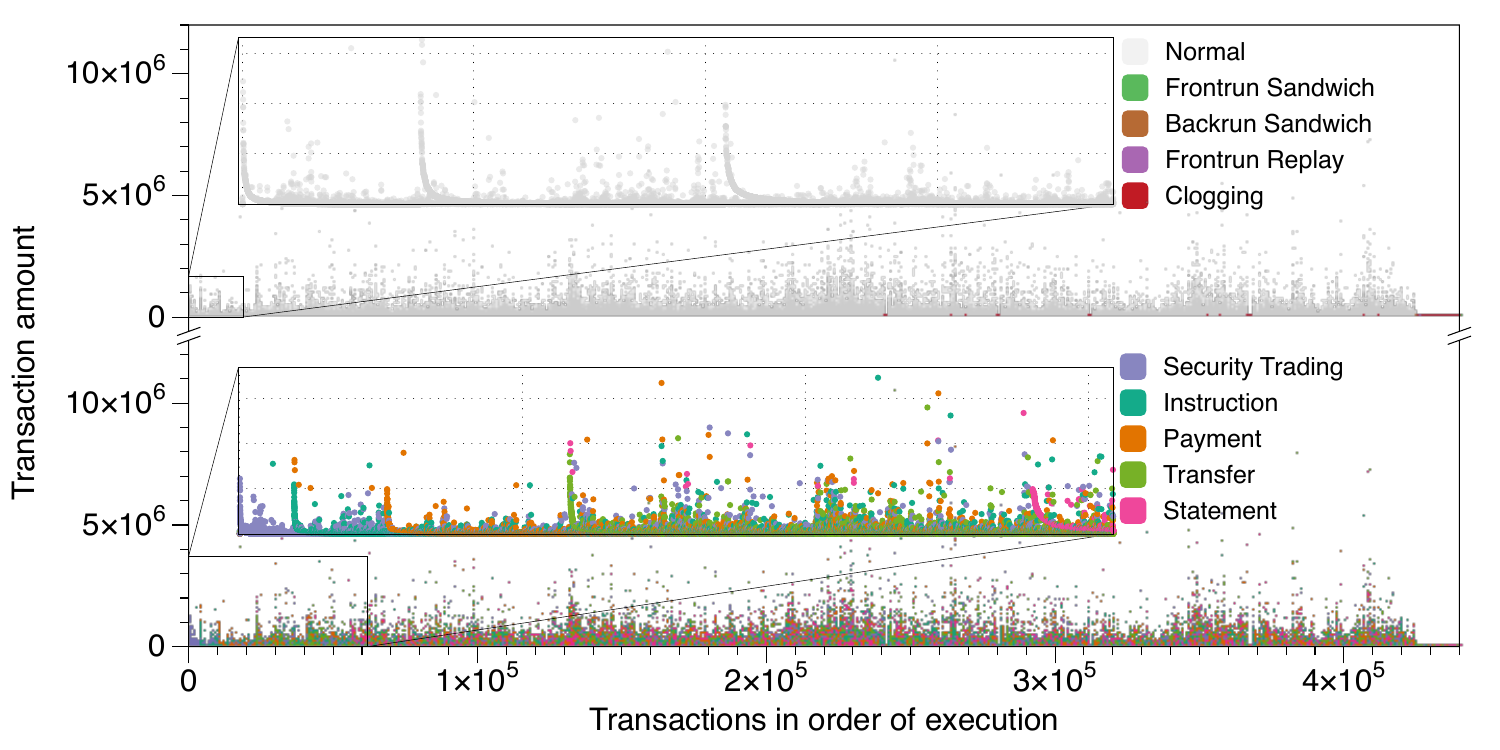}
    \caption{Illustration of the relationship between block size and the probability of multiple conflicting blocks being mined during block propagation (Probability of forking).}
    \label{TODCTransactions2}
\end{figure*}

Subsequently, utilizing the identical data set, we applied DTSS with optimized parameter settings for a second simulation experiment. As depicted in \reffig{TODCTransactions2}, the upper portion displays the distribution of adversarial transactions. Owing to a significantly reduced execution priority, the majority of adversarial transactions within the block are considerably diminished, with the exception of some Decoy transactions. These Decoy transactions, indistinguishable from regular transactions in the absence of specific preconditions, are thus screened by DTSS and retained. The lower half elucidates the distribution of various transaction types. A more explicit observation reveals that, in accordance with our established transaction execution priorities, Security Trading are preferentially executed, followed by Instructions, Payments, Transfers, and lastly, Statements.

\subsubsection{Quantitative Evaluation of DTSS Effectiveness}

Upon identifying variations in the execution priority within the transaction order, a systematic investigation was performed utilizing the Independent Samples T-Test, a statistical method that identifies the existence of any significant difference between the means of two independent groups. The Independent Samples T-Test is represented by the formula:

\begin{equation}
\displaystyle
\resizebox{0.24\hsize}{!}{
$t = \frac{M1 - M2}{\sqrt { \frac{s1^2}{n1} + \frac{s2^2}{n2} }}$
}
\end{equation}

Here, $M1$ and $M2$ stand for the sample means, $s1^2$ and $s2^2$ denote the sample variances, and $n1$ and $n2$ are the sample sizes. This test was deployed to compare the means of two independent groups on the same continuous, dependent variable - the execution priority. It facilitated the determination of whether there exists a statistically significant difference in the mean execution priority across various types of transactions. T-tests were conducted for each transaction type to evaluate the impact of DTSS on transaction execution priority. The p-values, \textcolor{black}{which can be found in t-distribution table corresponding to the t-values obtained}, were documented in TABLE \ref{t-test}. The analysis unveiled that DTSS significantly influenced the execution priority for each transaction type. Further, the degree of variation in the execution priority was computed using:

\begin{equation}
\displaystyle
\resizebox{.56\hsize}{!}{
$Shift = \frac{Mean(O_1) - Mean(O_2)}{Mean(O_2)}\times 100\%$
}
\end{equation}

Here, $O_1$ is the execution priority when DTSS is applied, and $O_2$ is the execution priority when DTSS is not applied. As depicted in TABLE \ref{t-test}, the Frontrun Sandwich, Backrun Sandwich transaction, Frontrun replay transaction, and Clogging transaction experienced shifts of 59.60\%, 55.94\%, 61.58\%, and 58.98\% respectively. In contrast, the normal transaction exhibited a decrease in shift by 6.24\%.

\begin{table*}
\footnotesize
\caption{\textcolor{black}{T-Test Result: DTSS Impact on Transaction Execution Priority}}
\label{t-test}
\centering
\resizebox{0.76\hsize}{!}{
  \begin{tabular}{c|c|c|c|c|c}
  \hline\hline
  \specialrule{0.00em}{3pt}{1pt} 
  \textbf{Transaction Types} & \textcolor{black}{\textbf{T-Value}}& \textbf{P-Value} & \textbf{$Mean(O_1)$} & \textbf{$Mean(O_2)$} & \textbf{Shift} \\
  \specialrule{0em}{3pt}{1pt}
  \hline
  \specialrule{0em}{3pt}{1pt}
  \textbf{Normal}   & 2362.49    & 0     & 210036.537174  &   223491.145347    & -6.24\%    \\
  \specialrule{0em}{3pt}{1pt}
  \hline
  \specialrule{0em}{3pt}{1pt}
  \textbf{Frontrun Sandwich}  & 1660.99 & $8.819144603221485\times 10^{-283}$       & 384989.292719 & 242157.379336 & 58.98\%             \\
  \specialrule{0em}{3pt}{1pt}
  \hline
  \specialrule{0em}{3pt}{1pt}
  \textbf{Backrun Sandwich}  & 1733.00     &  $9.196243840969524\times 10^{-293}$  &   386488.467742 & 242159.379336 &  59.60\%          \\
  \specialrule{0em}{3pt}{1pt}
  \hline
  \specialrule{0em}{3pt}{1pt}
  \textbf{Frontrun replay}  & 1760.72  & $7.186308877385129\times 10^{-296}$       & 385842.814925 & 238787.825170   &  61.58\%             \\
  \specialrule{0em}{3pt}{1pt}
  \hline
  \specialrule{0em}{3pt}{1pt}
  \textbf{Clogging}     & 24359.09      & 0   & 386273.315839 &  247698.844124  & 55.94\%             \\
  \specialrule{0em}{3pt}{1pt}
  \hline        
  \end{tabular}
}
\end{table*}

\subsubsection{Analysis \textcolor{black}{on} the Influence of Forking Threats on DTSS}

\begin{figure}[H]
    \makeatletter
    \def\@captype{figure} 
    \makeatother
    \centering
    \includegraphics[width=0.45 \textwidth]{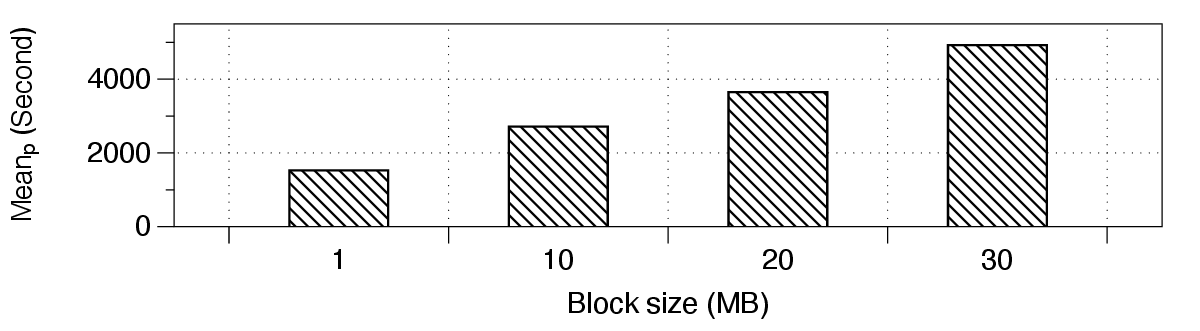}
    \caption{Diagram of the relationship between block size and the average block propagation time between nodes.}
    \label{BlockPropagation}
\end{figure}

\textcolor{black}{In this experiment, we establish a Blockchain network consisting of 6000 nodes by Simblock \cite{8751431}}. \textcolor{black}{To evaluate the time taken by the propagation of blocks of varying sizes within the Blockchain network, we consider the geographical distribution of nodes, as well as the allocation of network bandwidth and delay between different regions, as explained in  Table~\ref{BandwidthPram}. Next}, we simulate the Mean Propagation Duration (denoted as $Mean_p$) between two distinct nodes for blocks of diverse sizes, specifically 1MB, 10MB, 20MB, and 30MB. Given that each node exhibits a certain degree of randomness within the set probability range, the value of the variable $Mean_p$ yet can be expressed through \refeqs{BlockPropagation2}. As shown in \reffig{BlockPropagation}, the $Mean_p$ reveals a variation in response to the block sizes, suggesting that larger blocks necessitate a longer propagation time through the Blockchain network to attain consensus.

\begin{table}[H]
\caption{\textcolor{black}{Default network settings in SimBlock}}
  \centering
\resizebox{0.82\hsize}{!}{
\begin{threeparttable}
\footnotesize
\begin{tabular}{p{0.6cm}<{\centering} |p{8cm}<{\centering} }
\hline\hline
\specialrule{0em}{3pt}{1pt}
A1 & \tabincell{c}{R1: 33.16\%, R2: 49.98\%, R3: 0.09\%, R4: 11.77\%, R5: 2.24\%, R6: 1.95\%} \\
\specialrule{0em}{3pt}{1pt}
\hline
\specialrule{0em}{3pt}{1pt}
A2 & 
$\begin{array}{@{}r@{}c@{}c@{}c@{}c@{}c@{}c@{}c@{}}
   & R1 & R2 & R3 & R4 & R5 & R6  \\
   \left.\begin{array}{c} R1 \\R2 \\R3 \\R4 \\R5 \\R6 \end{array}
   \right(
        & \begin{array}{c} 2.29 \\ 2.29 \\ 0.69\\ 1.87 \\ 1.22 \\ 1.35 \end{array}
        & \begin{array}{c} 2.29 \\ 2.47 \\ 0.69 \\ 1.87 \\ 1.22 \\ 1.35 \end{array}
        & \begin{array}{c} 0.69 \\ 0.69 \\ 0.69 \\ 0.69 \\ 0.69 \\ 0.69 \end{array}
        & \begin{array}{c} 1.87 \\ 1.87 \\ 0.69 \\ 1.87 \\ 1.22\\ 1.35 \end{array}
        & \begin{array}{c} 1.22 \\ 1.22 \\ 0.69 \\ 1.22 \\ 1.22\\ 1.22 \end{array}
        & \begin{array}{c} 1.35 \\ 1.35 \\ 0.69 \\ 1.35\\ 1.22 \\ 1.35 \end{array}
        & \left)\begin{array}{c} \\ \\ \\ \\ \\ \\ \end{array}\right.
\end{array}$
\\
\specialrule{0em}{3pt}{1pt}
\hline
\specialrule{0em}{3pt}{1pt}
A3 & 
$\begin{array}{@{}r@{}c@{}c@{}c@{}c@{}c@{}c@{}c@{}}
   & R1 & R2 & R3 & R4 & R5 & R6  \\
   \left.\begin{array}{c} R1 \\R2 \\R3 \\R4 \\R5 \\R6 \end{array}
   \right(
        & \begin{array}{c} 32 \\ 124 \\ 184\\ 198 \\ 151 \\ 189 \end{array}
        & \begin{array}{c} 124 \\ 11 \\ 227 \\ 237 \\ 252 \\ 294 \end{array}
        & \begin{array}{c} 184 \\ 227 \\ 88 \\ 325 \\ 301 \\ 322 \end{array}
        & \begin{array}{c} 198 \\ 237 \\ 325 \\ 85 \\ 58\\ 198 \end{array}
        & \begin{array}{c} 151 \\ 252 \\ 301 \\ 58 \\ 12\\ 126 \end{array}
        & \begin{array}{c} 189 \\ 294 \\ 322 \\ 198\\ 126 \\ 16 \end{array}
        & \left)\begin{array}{c} \\ \\ \\ \\ \\ \\ \end{array}\right.
\end{array}$\\
\specialrule{0em}{3pt}{1pt}
\toprule
\end{tabular}
\begin{tablenotes}
\footnotesize
\item A1: Node Geographical Regions Distribution Percentage $Reg(N)$; A2: Network Bandwidth (Mbps) Between Each Pair of Regions $BAND(from,to)$; A3: Average Network Delay (Millisecond) Between Each Pair of Regions $DELAY(from,to)$
\item R1: North America; R2: Europe; R3: South American; R4: Asia Pacific; R5: Japan; R6: Australia;
\end{tablenotes}
\end{threeparttable}
}
\label{BandwidthPram}
\end{table}

\begin{footnotesize}
    \begin{equation}
        Mean_p =  \displaystyle\frac{\sum_{n=1}^{6000-1}(\frac{Size(B)}{Bandwidth(N_{f},N_{t})} + Delay(N_{f},N_{t}))}{6000}
        \label{BlockPropagation2}
    \end{equation}
\end{footnotesize}

\begin{figure}[H]
    \makeatletter
    \def\@captype{figure} 
    \makeatother
    \centering
    \includegraphics[width=0.4 \textwidth]{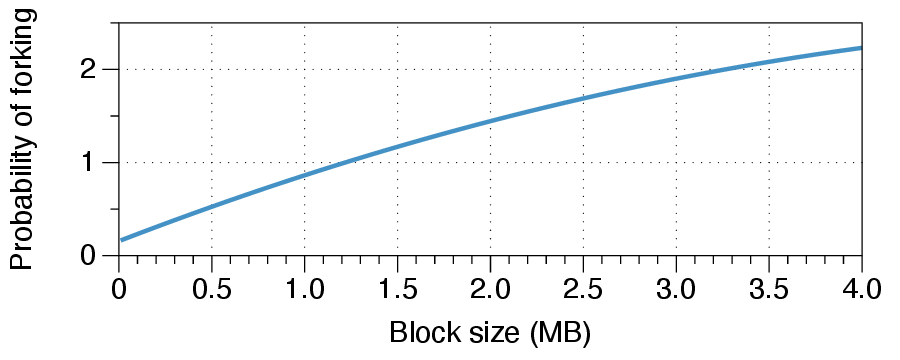}
    \caption{Illustration of the relationship between block size and the probability of multiple conflicting blocks being mined during block propagation (Probability of forking).}
    \label{ForkingRate}
\end{figure}

The probability of generating another conflicting block during block transmission, under different block sizes, was estimated according to \refeqs{ForkRatio}. The results in \reffig{ForkingRate} \textcolor{black}{show} the relationship between block size and the probability of forking.

\begin{figure}[H]
    \makeatletter
    \def\@captype{figure} 
    \makeatother
    \centering
    \includegraphics[width=0.45 \textwidth]{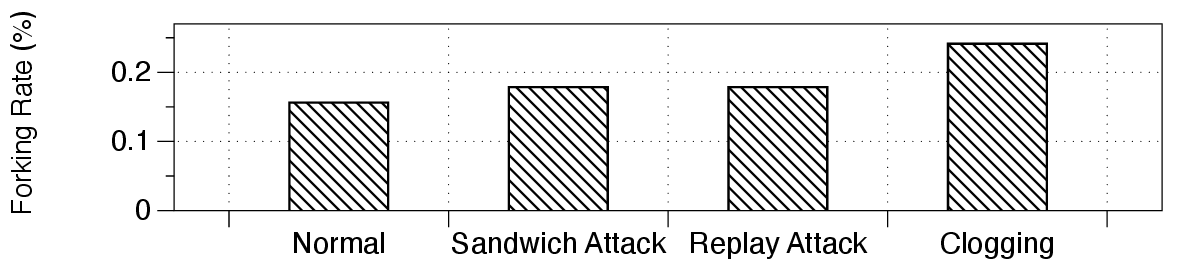}
    \caption{Fork rate of block attacked by Sandwich attack, replay attack or clogging. And fork rate of normal block}
    \label{BlockForkingRate}
\end{figure}

For miners who do not adhere to DTSS, under the parameter settings obtained through GA optimization, the block sizes of blocks containing Clogging, Sandwich attacks, and replay attacks are determined as 0.1485MB, 0.0265MB, and 0.0265MB, respectively. On the other hand, the block size of a normal block is 0.013MB. By referring to \refeqs{ForkRatio}, \textcolor{black}{it can be observed in \reffig{BlockForkingRate} that} the fork rates for blocks containing Clogging, Sandwich attacks, and replay attacks are calculated as 0.2413\%, 0.1785\%, and 0.1785\%, respectively. In contrast, the fork rate for a normal block is determined to be 0.1562\%. Notably, a decrease in fork rate is observed in normal blocks, indicating that blocks containing adversarial transactions face a higher risk of being orphaned compared to normal blocks. This, in turn, serves as a deterrent against potential attacks targeting the victim.

Miners who adhere to DTSS significantly reduce the likelihood of including adversarial transactions in their blocks. This is due to the fact that adversarial transactions are assigned scores that are substantially lower than those of normal transactions. By following the protocol and excluding transactions with low scores, normal miners can ensure that their blocks primarily consist of transactions with higher scores, which are indicative of their legitimacy and reliability. 


\subsection{Comparison with Existing Solutions}

Various mechanisms have been proposed to tackle inherent challenges related to transaction ordering and fairness. Each mechanism, unique in its approach and design, aspires to lessen risks associated with decentralized exchanges. Nevertheless, these mechanisms often face limitations when confronted with the intricate, ever-changing nature of permissionless environments and sophisticated attacks. We present a comprehensive comparison of three such models – Fair Ordering, Fair Ordering Extension, and Application-Specific Blockchain Extractable Value (BEV) Mitigation – and juxtapose them with DTSS, as detailed in Table \ref{tab:comparison}.

\begin{table*}
\footnotesize
\caption{Comparison of Transaction Ordering Mechanisms}
\label{tab:comparison}
\centering
\resizebox{0.86\hsize}{!}{
\begin{tabular}{c|c|c|c|c}
\hline\hline
  \specialrule{0em}{3pt}{1pt}
\textbf{Mechanism} & \textbf{Fair Ordering \textcolor{black}{\cite{10.1007/978-3-030-56877-1_16}}} & \textbf{Fair Ordering Extension \textcolor{black}{\cite{10.1145/3494105.3526239}}} & \textbf{BEV Mitigation \textcolor{black}{\cite{zhou2021a2mm}}} & \textbf{DTSS} \\
\specialrule{0em}{3pt}{1pt}
\hline
  \specialrule{0em}{3pt}{1pt}
Permissionless Environment Compatibility & Limited & Yes & Yes & Yes \\
\specialrule{0em}{3pt}{1pt}
\hline
  \specialrule{0em}{3pt}{1pt}
Resilience to Order Manipulation Attacks & No & Yes & Yes & Yes \\
\specialrule{0em}{3pt}{1pt}
\hline
  \specialrule{0em}{3pt}{1pt}
Requires Additional Coordination & No & Yes & Yes & No \\
\specialrule{0em}{3pt}{1pt}
\hline
  \specialrule{0em}{3pt}{1pt}
Flexible and Scalable & Limited & Yes & Limited & Yes \\
\specialrule{0em}{3pt}{1pt}
\hline
  \specialrule{0em}{3pt}{1pt}
Effective Against Diverse Attacks & No & Yes & Limited & Yes \\
\specialrule{0em}{3pt}{1pt}
\hline
  \specialrule{0em}{3pt}{1pt}
Integrated within Consensus Protocol & No & No & No & Yes \\
\specialrule{0em}{3pt}{1pt}
\hline
  \specialrule{0em}{3pt}{1pt}
Adaptable Transaction Ordering & No & No & No & Yes \\
\specialrule{0em}{3pt}{1pt}
\hline
\end{tabular}
}
\end{table*}

\textcolor[RGB]{0,0,0}{The Fair Ordering model, while effective in traditional permissioned blockchain environments, struggles to scale in permissionless settings. Designed for a fixed set of trusted protocol nodes, it becomes vulnerable in scenarios where adversarial participants can exploit order manipulation. Its reliance on a static network structure limits its applicability in dynamic and decentralized ecosystems, making it less suitable for modern blockchain networks.}

\textcolor[RGB]{0,0,0}{The Fair Ordering Extension attempts to address these limitations by introducing mechanisms to enforce fairness in permissionless systems. However, the inclusion of an auxiliary finalization coordination mechanism significantly increases system complexity. This added layer places a computational and operational burden on miners, complicating their coordination within a distributed network. As a result, it risks delaying block finalization and reducing overall throughput, particularly in high-traffic blockchain environments.}

\textcolor[RGB]{0,0,0}{The Application-Specific BEV Mitigation mechanism demonstrates efficacy in specific contexts, such as Automated Market Maker (AMM) exchanges, by mitigating BEV-related attacks like arbitrage. However, its scope is inherently narrow, as it relies on efficient communication across multiple AMM exchanges. This restricts its ability to handle a broader range of BEV attacks, including non-AMM-associated threats like clogging or transaction replay. Its lack of scalability and flexibility renders it insufficient for addressing the diverse vulnerabilities present in complex blockchain ecosystems.}

\textcolor[RGB]{0,0,0}{In contrast, the Distributed Transaction Sequencing Strategy (DTSS) offers a comprehensive and adaptable solution. By integrating directly into the consensus protocol, DTSS eliminates the need for additional communication layers or auxiliary coordination mechanisms. This streamlines the role of miners, ensuring that transaction ordering fairness is enforced natively during block creation. Moreover, DTSS's dynamic adaptability allows it to adjust transaction prioritization based on multiple attributes, such as transaction type, fee, and initiation time. This ensures a versatile and robust defense against a wide range of BEV attacks, including those beyond the scope of other mechanisms, while maintaining efficiency and scalability across diverse blockchain environments.}

\subsection{\textcolor[RGB]{0,0,0}{Scalability}}

\textcolor[RGB]{0,0,0}{DTSS not only plays a crucial role in controlling how transactions are incorporated into a block but is also highly applicable to various types of Blockchain scalability solutions, both at the first layer and the second layer.}

\textcolor[RGB]{0,0,0}{First layer scalability solutions aim to enhance the Blockchain's throughput by incorporating more transactions into each block \cite{KUZNETSOV2024101315}. Several prominent examples include Xtreme Thinblocks (XThin) \cite{Thinblocks}, Compact Blocks \cite{Compactblock}, and Graphene \cite{ozisik2019graphene}. XThin focuses on reducing the size of blocks transmitted between nodes by sending only the missing transactions, thus improving block propagation time. Compact Blocks similarly reduce block size by sharing only the necessary transaction data between nodes, enabling faster block dissemination. Graphene employs an efficient block propagation protocol that uses Bloom filters and Invertible Bloom Lookup Tables (IBLTs) to minimize the data exchanged between nodes. DTSS is particularly effective in these first-layer solutions as they provide more leaf nodes for the DTSS to allocate transactions.}

\textcolor[RGB]{0,0,0}{Second layer scalability solutions, such as the Lightning Network \cite{poon2016bitcoin}, aim to facilitate off-chain transactions, thereby reducing the load on the main Blockchain. The Lightning Network allows for the creation of payment channels where multiple transactions can occur off-chain, with only the opening and closing of the channels being recorded on the Blockchain. DTSS is also applicable to second layer scalability solutions because the Lightning Network does not alter the transaction incorporation mechanism of the block. By employing DTSS, the submission and sequencing of transactions that eventually get recorded on-chain (when channels are opened or closed) can be managed deterministically. This ensures that the ordering of these transactions remains consistent and tamper-proof, maintaining the integrity and security of the Blockchain.}

\section{Conclusion}

This paper provides a comprehensive exploration of Blockchain technology, its applications in decentralized finance (DeFi), and the intricacies associated with transaction ordering dependency. We initiate our discussion with an in-depth analysis of various Blockchain activities, encompassing transaction processing, the role of miners, and the potential risks posed by extractable value.

We proceed to elucidate the emergent entities within the financial sector that have materialized owing to DeFi, and the possible benefits and challenges allied with the adoption of DeFi. Despite existing challenges, such as regulatory uncertainty and potential security vulnerabilities, the growth trajectory of DeFi seems relentless, indicating its potential to revolutionize traditional financial models.

A salient part of our exploration revolves around the threat posed by Blockchain Extractable Value (BEV) to transactions within the DeFi ecosystem, particularly in relation to Sandwich attacks. We underscore the paramount role of miners and the influence they wield in determining transaction order within blocks, leading to potential transaction-ordering dependence issues.

To mitigate these challenges, we introduce a novel Distributed Transaction Sequencing Strategy (DTSS). This strategy exploits the inherent forking threat in the Blockchain network, which is a function of the cumulative computational exertions of miners to validate and incorporate blocks into the Blockchain. Blocks of smaller size possess an advantage as they can achieve consensus more expeditiously, facilitating faster validation and integration into the Blockchain.

Our proposed mechanism discerns crucial transaction attributes and employs the Analytic Hierarchy Process (AHP) to assign weights to these attributes, thereby establishing transaction priorities. Through rigorous experimentation, we have tested various optimization algorithms and identified optimal parameter settings for DTSS.

\textcolor[RGB]{0,0,0}{The experimental findings demonstrate that DTSS can efficaciously enforce miners to comply with predefined transaction ordering rules, thereby addressing transaction-ordering dependence challenges, enhancing trust among network participants. DTSS offers consistent, tamper-proof transaction ordering, benefiting high-security and real-time applications such as DeFi, traditional banking, and security trading systems, improving transaction management across diverse industries.}

In our proposed DTSS, we utilize the concept of forking as ``The Sword of Dharma''. Miners are impelled to adhere to these rules, ensuring the integrity of the transaction process. Upon non-compliance, DTSS enforces a penalty whereby the offending miner's block will be forked by other miners. This establishes a distributed self-regulating system where rule adherence is incentivized, effectively addressing BEV concerns within the Blockchain ecosystem.

Kelkar in \cite{10.1145/3494105.3526239}, posited that ``We find that permissionless order-fairness is impossible to achieve and not particularly interesting in the presence of completely dynamic adversaries which can corrupt and kill nodes before blocks are mined.'' While DTSS is primarily proposed for consortium Blockchains, we postulate its potential to counter BEV threats in permissionless settings as well. This conjecture, grounded in the inherent robustness of DTSS, necessitates further empirical validation and will form the crux of our future research endeavors.

\textcolor[RGB]{0,0,0}{Despite DTSS's strengths, it has a notable shortcoming in addressing arbitrage activities. Arbitrage is a strategic practice that involves the simultaneous purchase and sale of assets in different markets to capitalize on price discrepancies. By exploiting these disparities, Blockchain arbitrageurs aim to secure assets at lower prices and sell them at higher prices, thereby aligning market prices and enhancing liquidity. Since arbitrage does not necessarily depend on the order of transaction execution but rather on price differences across markets, DTSS’s mechanism of ensuring chronological order does not address the fundamental nature of arbitrage. We will strive to develop solutions for this issue in future work.}



\textcolor[RGB]{0,0,0}{Further extensions of this work also include developing enhanced security mechanisms to protect against emerging threats and ensuring interoperability across different Blockchain platforms. Optimizing DTSS for specific use cases such as decentralized finance (DeFi) and traditional banking systems is also a key area for future research. While DTSS does not enhance scalability on its own, further study is needed to combine it with scalability solutions for efficient high transaction volume management. Leveraging machine learning techniques to predict transaction patterns and optimize sequencing could further enhance the adaptability and intelligence of the DTSS system. Additionally, we plan to explore the integration of intelligent data protocols for the edge \cite{9748997} \cite{10265750}, which could improve the efficiency and robustness of data exchanges within the DTSS framework.}

\section*{Acknowledgments}
This research was funded by the University of Macau (file no. MYRG2022-00162-FST and MYRG2019-00136-FST).

\section*{CRediT author statement}
Xiongfei Zhao: Conceptualization, Methodology, Software, Writing-Original draft preparation. Hou-Wan Long: Methodology, Software, Writing-Original draft preparation. Zhengzhe Li: Methodology, Software, Writing-Original draft preparation. Jiangchuan Liu: Conceptualization, Methodology, Supervision, Writing-Reviewing. Yain-Whar Si: Supervision, Writing-Reviewing and Editing, Funding acquisition.









\end{document}